\documentclass[a4paper,10pt,twoside]{cpc-hepnp}
\usepackage{multicol}
\usepackage{graphicx}
\usepackage{epstopdf,epsfig,subfigure}
\usepackage{booktabs}
\usepackage{amssymb,bm,mathrsfs,bbm,amscd}
\usepackage[tbtags]{amsmath}
\usepackage{lastpage}
\usepackage{CJK}
\usepackage{colortbl}
\definecolor{blackblue}{RGB}{25,25,120}
\definecolor{deepblue}{RGB}{0,0,51}
\usepackage[colorlinks,linkcolor=blackblue,anchorcolor=blackblue,citecolor=blackblue,urlcolor=blackblue]{hyperref}

\begin{document}
\begin{CJK*}{GBK}{song}



\title{Deep learning based track reconstruction on CEPC luminometer\thanks{supported by Joint Large-Scale Scientific Facility Funds of the NSFC and CAS, U1632104.}}

\author{%
      YANG Liu$^{1}$
\quad CAI Hao$^{1}$\email{hcai@whu.edu.cn}
\quad ZHU Kai$^{2}$\email{zhuk@ihep.ac.cn}%
}
%

\maketitle

\address{%
$^1$ School of Physics and Technology, Wuhan University, Wuhan 430072, China\\
$^2$ Institute of High Energy Physics, Chinese Academy of Sciences, Beijing 100049, China\\
}

\begin{abstract}
We study the track reconstruction algorithms of the CEPC luminometer. Depend on the
current geometry design, the conventional track reconstruction method is applied,
but it suffers the energy leakage problem when tracks
falling into the tile gaps regions. To solve this problem, a novel
reconstruction method based on deep neural networks has been investigated, and the reconstruction efficiency has been improved significantly, as well as the energy and direction resolutions.
This new reconstruction method is proposed to replace the conventional one for the CEPC
luminometer.
\end{abstract}




\begin{multicols}{2}

\section{Introduction}
The future Circular Electron Positron Collider (CEPC), which is proposed by Chinese high energy
physics community, will run as a Higgs factory at a center-of-mass energy of $240~\mathrm {GeV}$ to precisely study the properties of Higgs boson and
also be operated at Z pole,  ZH threshold, WW threshold, and Z line shape to precisely
measure the W and Z boson masses, widths, and couplings~\cite{CEPC_PreCDR}.
The physics goals of
CEPC require that the relative uncertainty of measured luminosity is smaller
than $10^{-3}$. The precision can not be reached by the measurements via main detectors of CEPC,
so a specific luminosity calorimeter is required.


The luminosity is determined by counting the small angle Bhabha
events with
\begin{equation}\label{Bhabha_lumi}
\mathcal{L} = \frac{N}{\epsilon\sigma(e^+e^- \to e^+e^-)},
\end{equation}
where $\mathcal{L}$ is the luminosity to be determined; $N$, $\epsilon$, $\sigma$ are the number of signal
events, detection efficiency, and the cross section of
Bhabha process, respectively. The cross section is given by
\begin{equation}\label{sigma_eq}
\sigma = \frac{16\pi\alpha^2}{S}(\frac{1}{\theta_{min}^2}-\frac{1}{\theta_{max}^2})\simeq\frac{16\pi\alpha^2}{S\theta_{min}^2}.
\end{equation}
Here $\theta_{min}$ and $\theta_{max}$ are the luminometer inner
and outer fiducial polar angles, respectively; $S$ is the square of center-mass-energy of Bhabha
event; $\alpha$ is the fine-structure constant.
This formula tells that the precision of cross section of Bhabha process heavily relies on the direction reconstruction, especially
the polar angle determination, of the incident track. For small angle, a small bias of polar angle
determination leads to larger variation on the luminosity determination, which is
\begin{equation}\label{L_theta}
\frac{\delta\mathcal{L}}{\mathcal{L}} = \frac{2\delta\theta}{\theta_{min}}.
\end{equation}
Here $\delta\mathcal{L}$ and $\delta\theta$ are the uncertainty of luminosity and the uncertainty of the measured polar angle, respectively.

\begin{center}
\includegraphics[width=6cm]{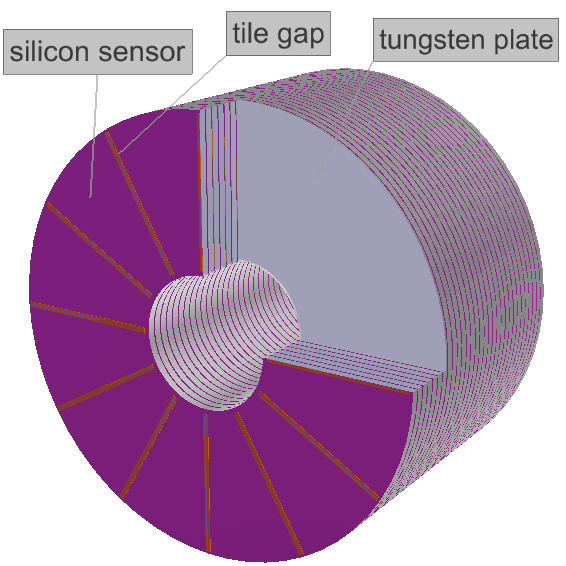}
\figcaption{\label{fig_lum_geo} (Color online) Sketch of mechanical structure of luminometer.}
\end{center}

The integrated luminosity calorimeter will be at the very forward region of CEPC.
At present, it is designed as a sandwich-type silicon-tungsten calorimeter,
which is similar to the future International
Linear Collider's (ILC) sub-detector LumiCal~\cite{Lcal_geo}. The luminometer  has two modules separately positioned at two sides of the interaction point along the beam axis
in $970~\mathrm{mm}$ away, and has a polar angle coverage
from $30~\mathrm{mrad}$ to $100~\mathrm{mrad}$.
Each module consists of 30 layers of tungsten with 1 radiation length thickness.
Each tungsten layer is followed by segmented silicon sensors planes. Radial outside space of the silicon/tungsten region is the electric readout. There are thin gaps between silicon/tungsten layers. Each sensor plane
has $48$ divisions and $64$ divisions in azimuthal and radial direction, respectively. Mechanically, only four sectors can
fit on a single tile of silicon, so there is an uninstrumented tie gap of 1.2 mm width between every four sectors. To compensate for the energy
loss of electrons transporting through these gaps, every other layer is designed to be rotated by $3.75^\circ$.

Accurate identification of Bhabha events depends in part on matching the energies, polar angle and
azimuthal angle of particles that
strike the opposing luminometer modules. Therefore, the reconstruction of the
incident track direction
and energy plays a very important role in the luminosity measurement.

Many successful applications of deep neural network (DNN)
are established in various fields (a general introduction of DNN can
be found in Ref.~\cite{ref_DNN}), including some exciting works in experimental high energy
physics, such as, track reconstruction in complex detector~\cite{DNN_work1},
discriminating signal events from background
events~\cite{DNN_work2,DNN_work3,DNN_work4,DNN_work5,DNN_work6} and simulations of particle collisions~\cite{DNN_work7}.
In this paper, we first introduce the track reconstruction result which is
based on the conventional methods.
Then, to deal with the tracks falling into tile gaps, which will result in lower efficiency or worse
resolutions with the conventional method, we investigate a novel method based on the DNN.
It turns out the DNN based method can solve this problem and provide higher
reconstruction efficiency and better resolutions in both energy and direction
than the conventional method.

\section{Monte Carlo simulation}

A {\sc Geant4}~\cite{geant4} application {\sc Mokka}~\cite{mokka} is
used to implement the detector geometry and supply
a valid way to simulate the physics processes of incident tracks transporting through
the luminometer and calculate the energy deposition in each sensor cell.
The parameters of the
luminometer used in simulation are listed in Table.~\ref{geo-param}.

To study the performance of the luminometer,
we have generated a Monte Carlo sample with $3\times10^5$ single positron track events,
which is used as the training set for the DNN based track reconstruction method.
The energy, polar angle ($\theta$), and azimuthal angle ($\phi$) of positrons distribute uniformly
between $(20, 200)~\mathrm{GeV}$, $(35, 90)~\mathrm{mrad}$,
and $(0, 2\pi)~\mathrm{rad}$, respectively.
Another 10 Monte Carlo samples with track energies from 20 to 200 GeV with step 20 GeV,
each of which contains $1\times10^4$ events, are generated
as the test sets.
\begin{center}
\tabcaption{ \label{geo-param}  Geometric parameters used for simulation.}
\footnotesize
\begin{tabular*}{80mm}{c@{\extracolsep{\fill}}cc}
\toprule
Element             & Value                     & Unit  \\
\hline
planes/module       & $30$                        & -\\
tiles/plane         & $12$                        & -\\
sectors/tile        & $4$                         & -\\
cells/sector        & $64$                        & -\\
Length              & $130.0$                     & $\mathrm{mm}$\\
Position(z)         & $\pm 970$                  & $\mathrm{mm}$\\
inner radius        & $30$                      & $\mathrm{mm}$\\
outer radius        & $100$                      & $\mathrm{mm}$\\
layer gap           & $0.25$                    & $\mathrm{mm}$\\
silicon thickness   & $0.32$                    & $\mathrm{mm}$\\
support thickness   & $0.2$                     & $\mathrm{mm}$\\
tile gap            & $1.2$                     & $\mathrm{mm}$\\
tungsten thickness  & $3.5$                     & $\mathrm{mm}$\\
sensor phi rotate   & $3.75$                    & degree\\
\bottomrule
\end{tabular*}
\end{center}

\section{Hit clustering}
For a physical event, it is possible that multiple incident tracks shoot into one luminometer module simultaneously. The clustering method is aiming at grouping the hit cells into different clusters, each of which is generated from a primary incident track.
We adopted the clustering method used in the ILC group, that is composed by steps of
selection of shower-peak layers, shower-peak layer clustering,
and 3D global clustering. The details of
the clustering method can be found in the Ref.~\cite{Clustering_method},
and is implemented by us with a series of Marlin~\cite{Marlin} processors.
%
\section{The conventional track reconstruction algorithm}

\subsection{Energy reconstruction}
The conventional energy reconstruction for a sampling calorimeter works in a simple way.
Usually, the total deposition energy in the sensitive region is a proportion of the incident particle energy.
To reconstruct the original energy of one particle, a correction factor (CF)
is multiplied by the summation of the deposition energy in
the sensitive sensor cells. Generally, the CF is relative to the incident track  energy.
Due to the effect of the
tile gap, the total deposited energy in the luminometer with a track shotting nearby a tile gap is
obviously lower than other regions as shown in Fig.~\ref{gap_eff}.
\begin{center}
\includegraphics[width=8cm]{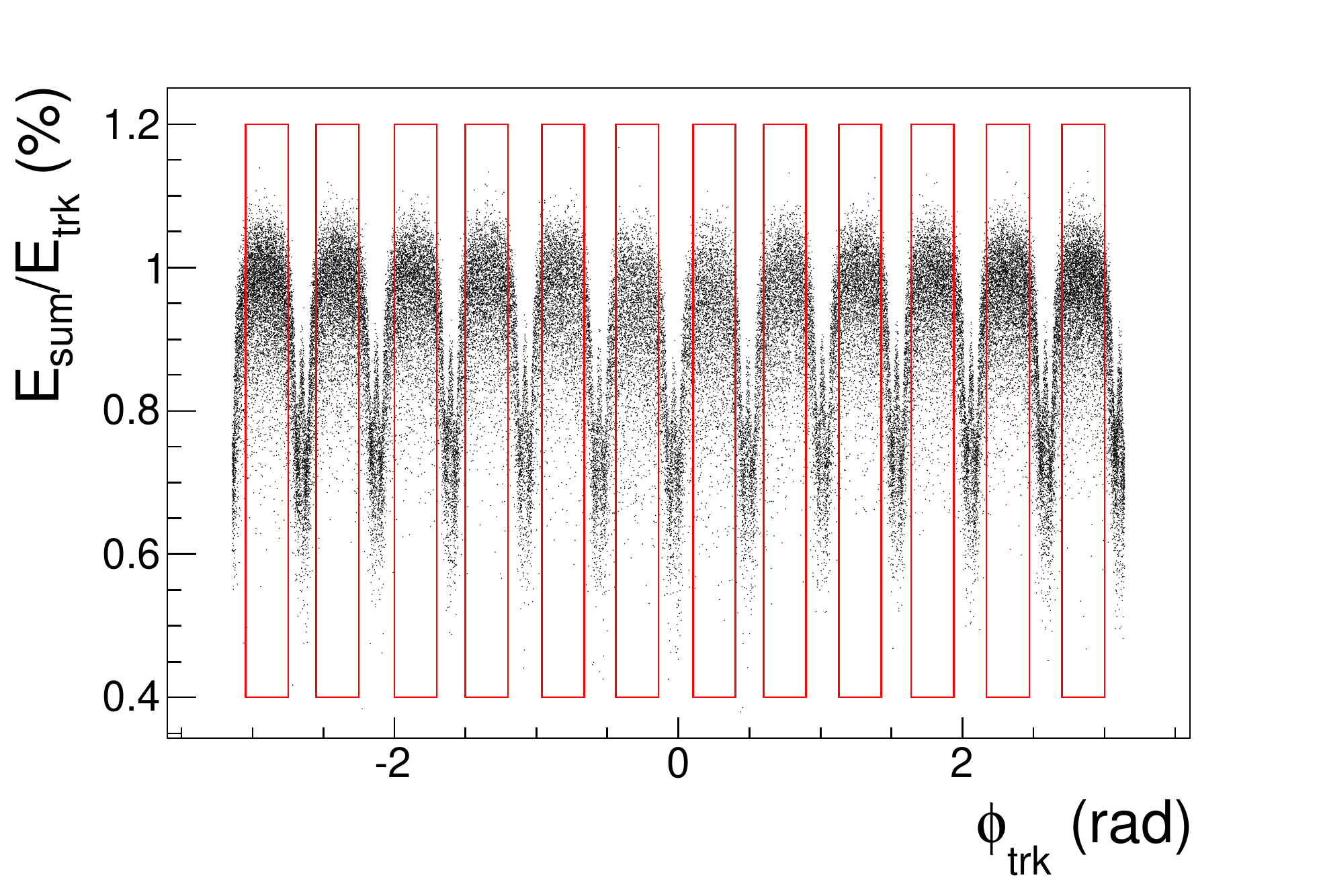}
\figcaption{\label{gap_eff} (Color online) The sum of deposited energy in sensor cells ($\rm{E_{sum}}$) divided by the origin track energy ($\rm{E_{trk}}$) versus the incident track azimuthal angle ($\mathrm{\phi_{trk}}$).
}
\end{center}
The tile gap effect will result in worse energy resolution.
To improve the resolution, we adopt the method that only take tracks shooting into non-gap  regions (region within 0.3~rad of
each tile center is chosen as the non-gap region).
Then we get the distributions of the $\rm{E_{sum}}/\rm{E_{trk}}$ as shown in Fig.~\ref{cut_gap},
where $\rm{E_{sum}}$ is the sum of the
deposited energy in the sensor cells and $\rm{E_{trk}}$ is the original energy of the track.
We fit the $\rm{E_{sum}}/\rm{E_{trk}}$ distributions with Gaussian functions to
extract the mean and width values,
where the inverse of the mean is taken as the CF and the width is taken as the energy resolution.
\begin{center}
{\includegraphics[width=4.5cm]{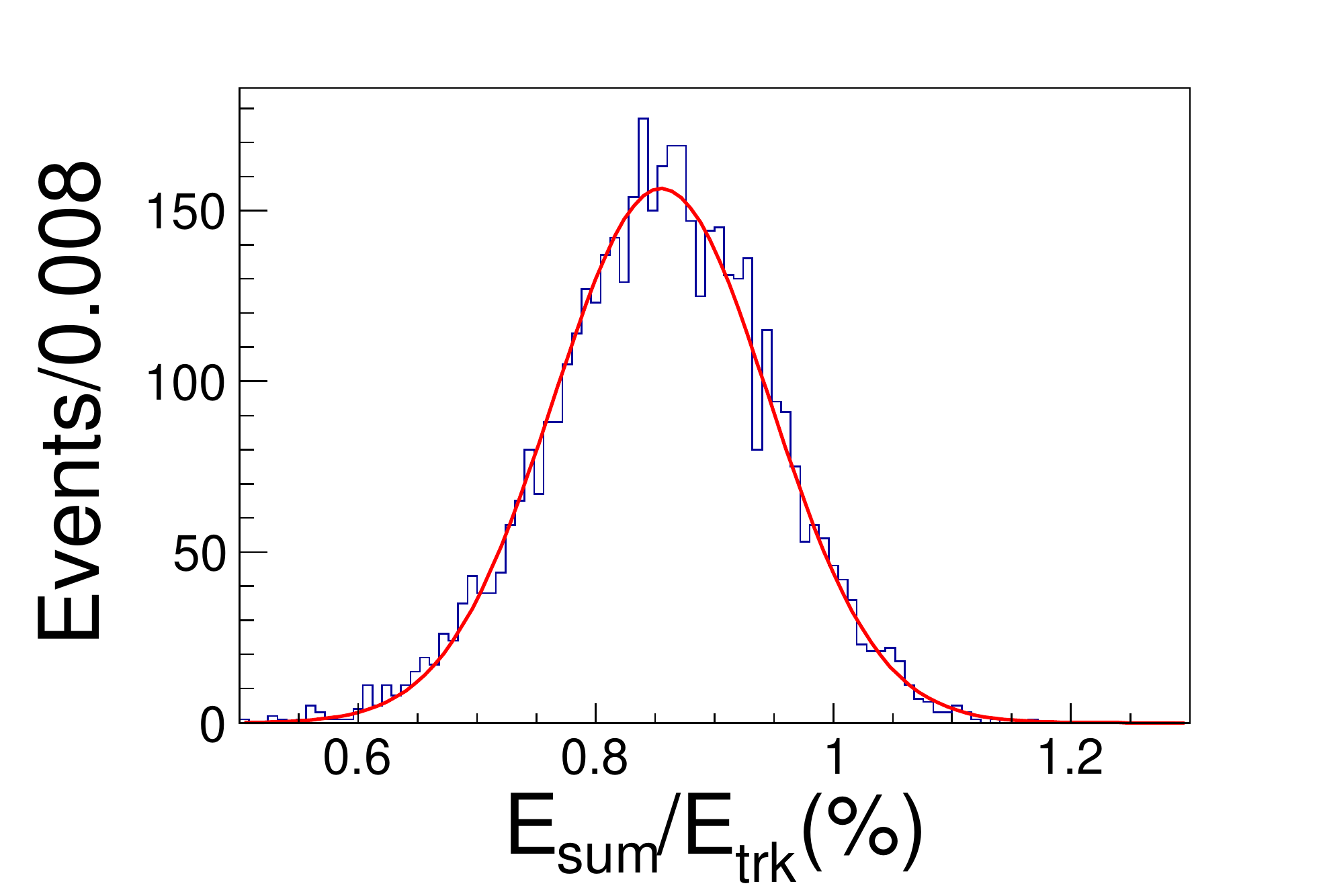}\put(-95,45){(${\rm E_{trk}}$ =20 GeV)}
\includegraphics[width=4.5cm]{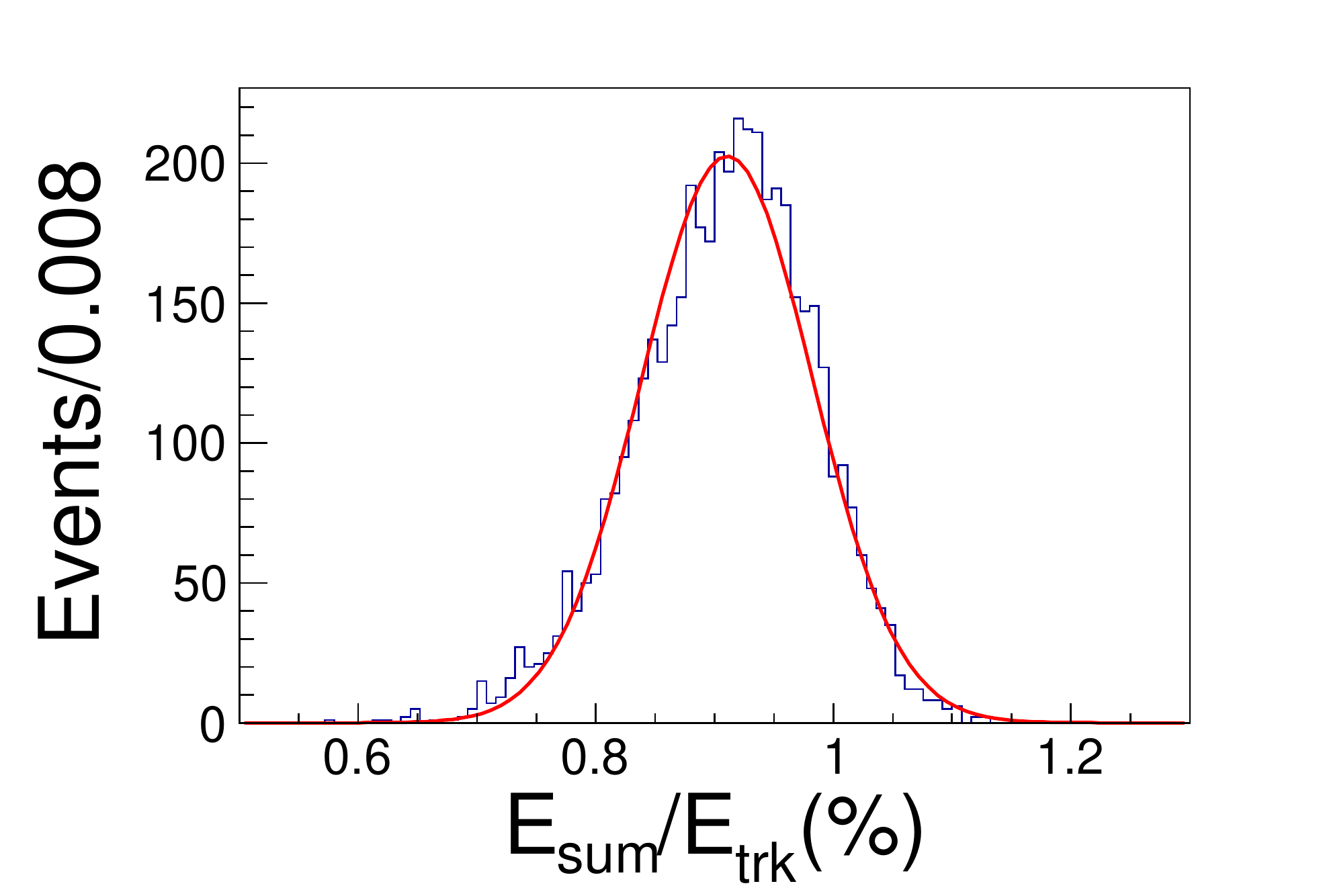}\put(-95,45){(${\rm E_{trk}}$ =40 GeV)}}
{\includegraphics[width=4.5cm]{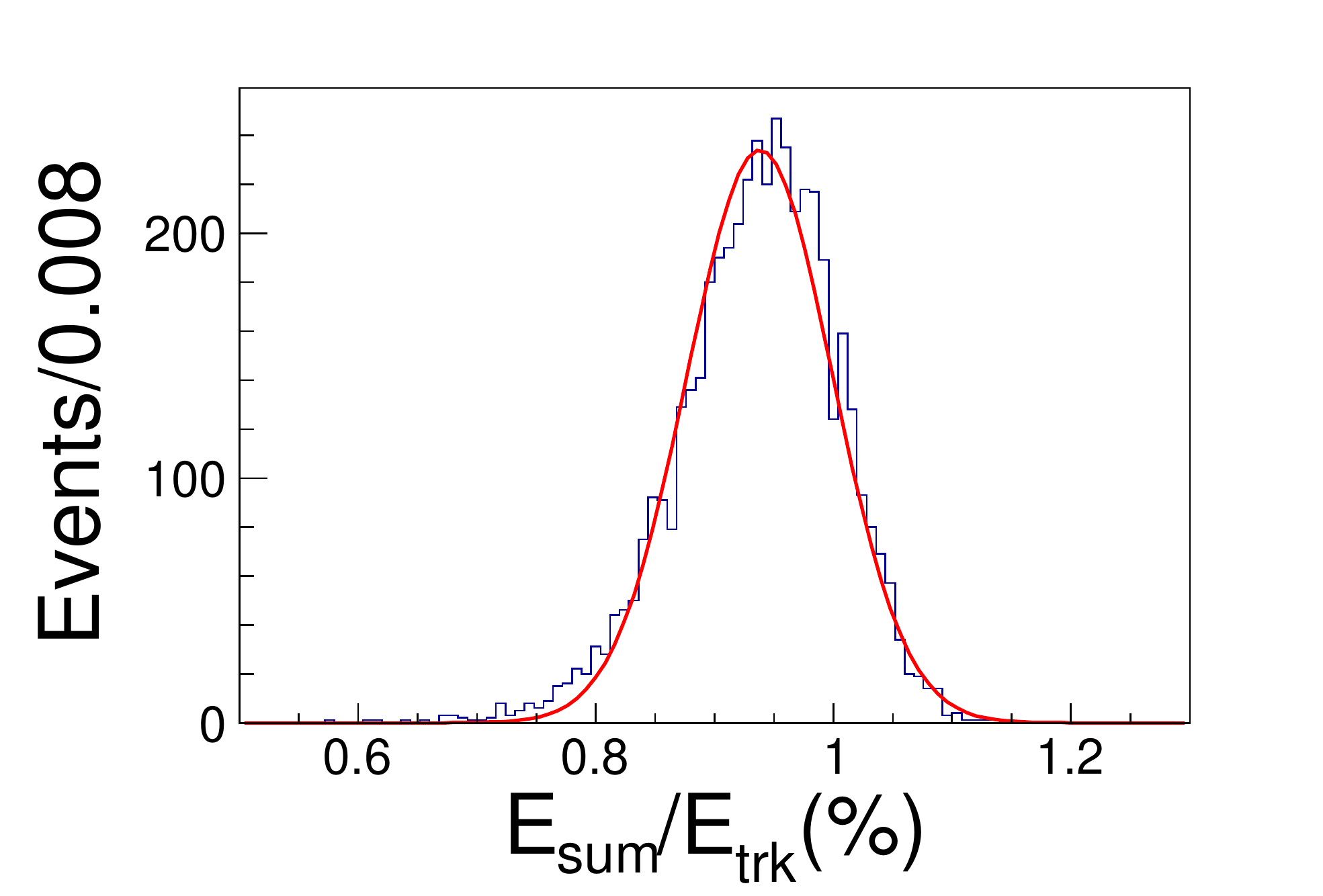}\put(-95,45){(${\rm E_{trk}}$ =60 GeV)}
\includegraphics[width=4.5cm]{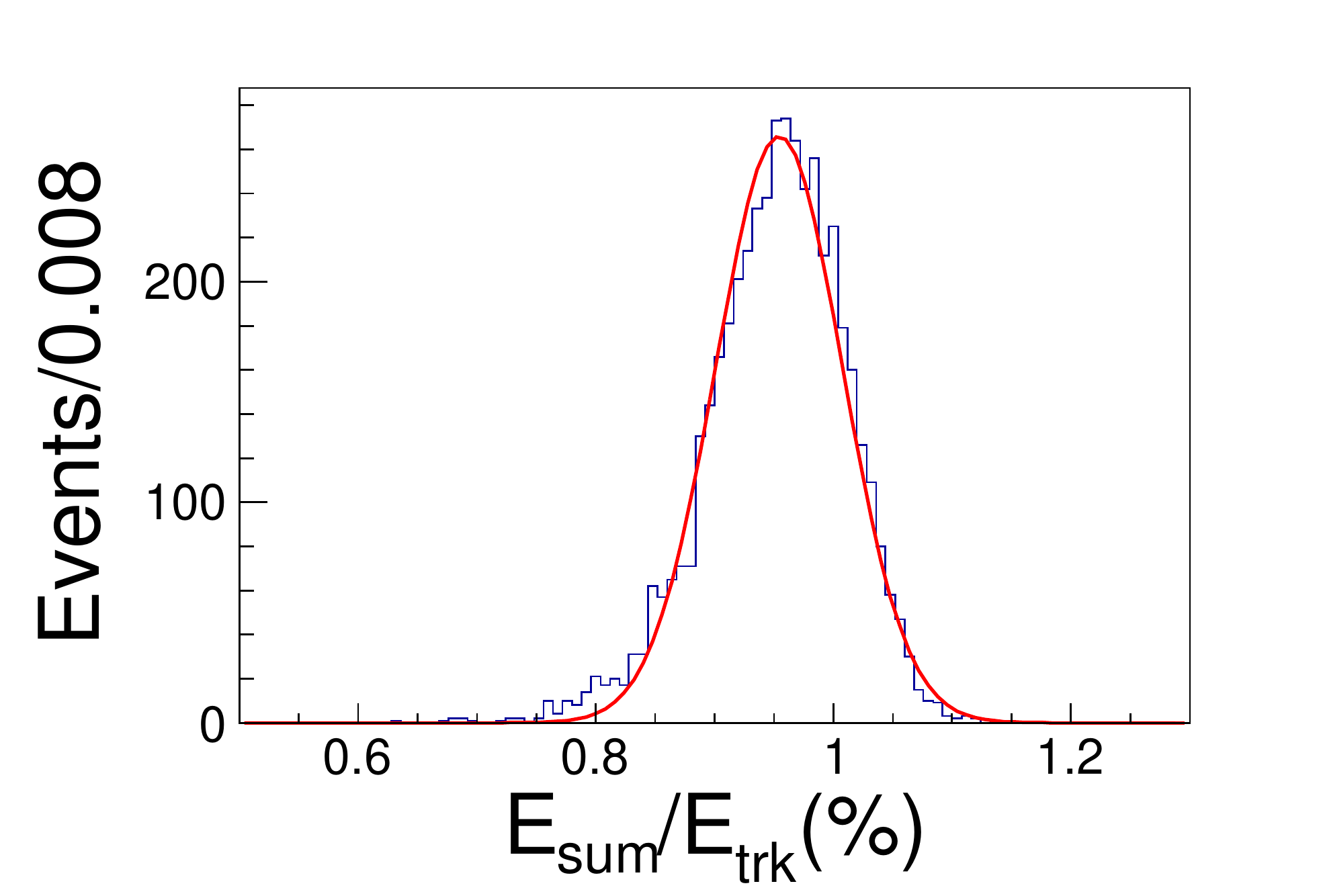}\put(-95,45){(${\rm E_{trk}}$ =80 GeV)}}
{\includegraphics[width=4.5cm]{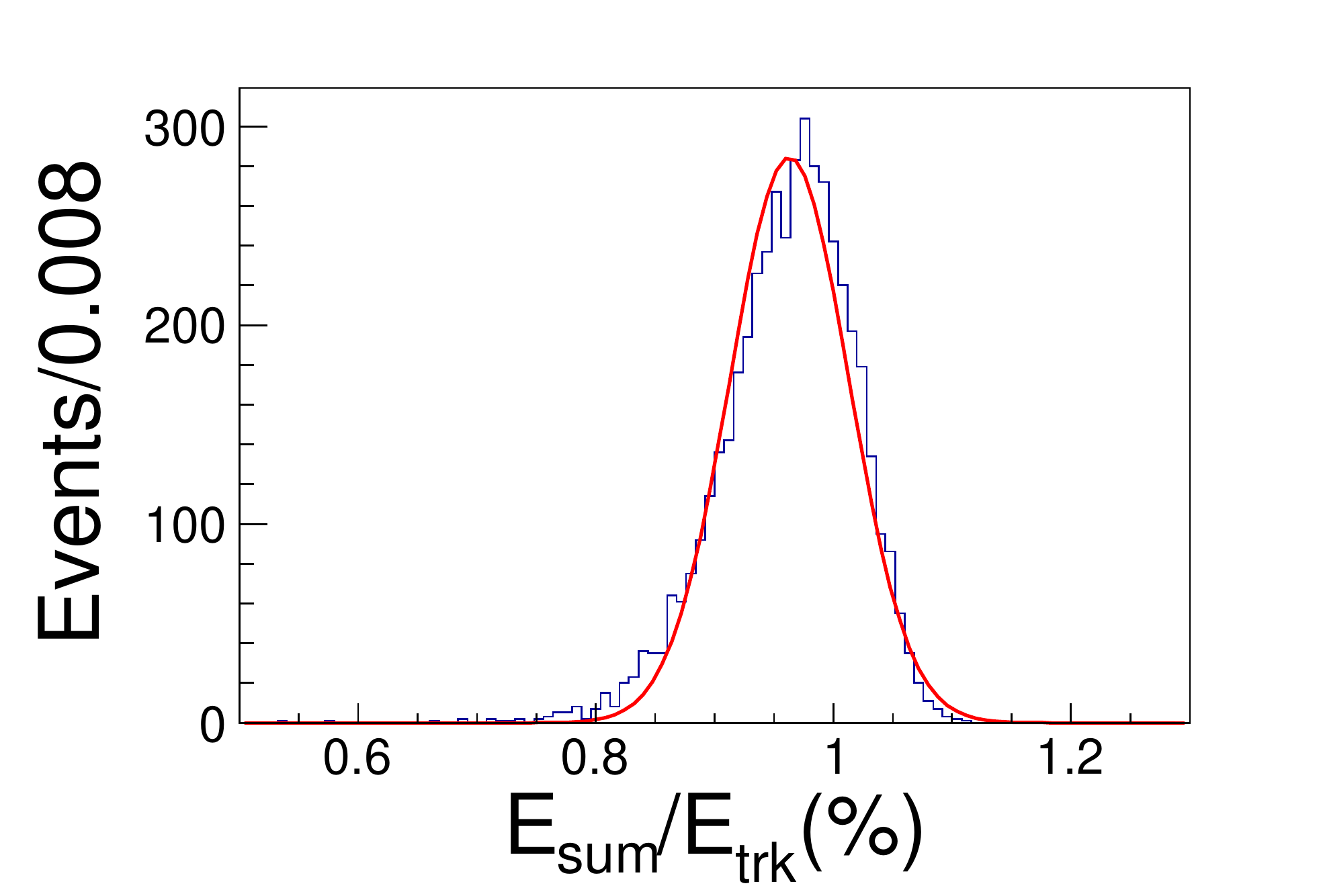}\put(-95,45){(${\rm E_{trk}}$ =100 GeV)}
\includegraphics[width=4.5cm]{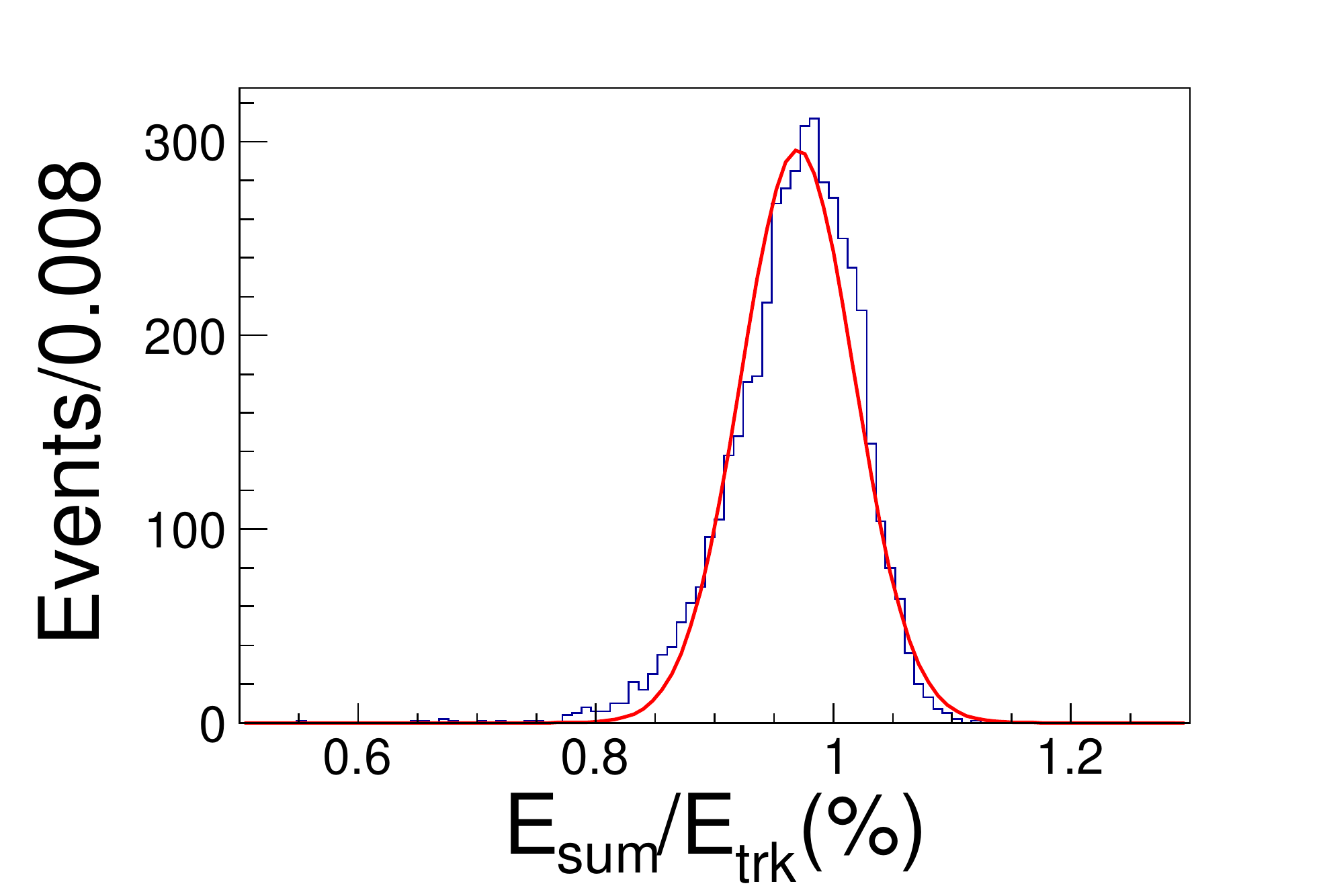}\put(-95,45){(${\rm E_{trk}}$ =120 GeV)}}
{\includegraphics[width=4.5cm]{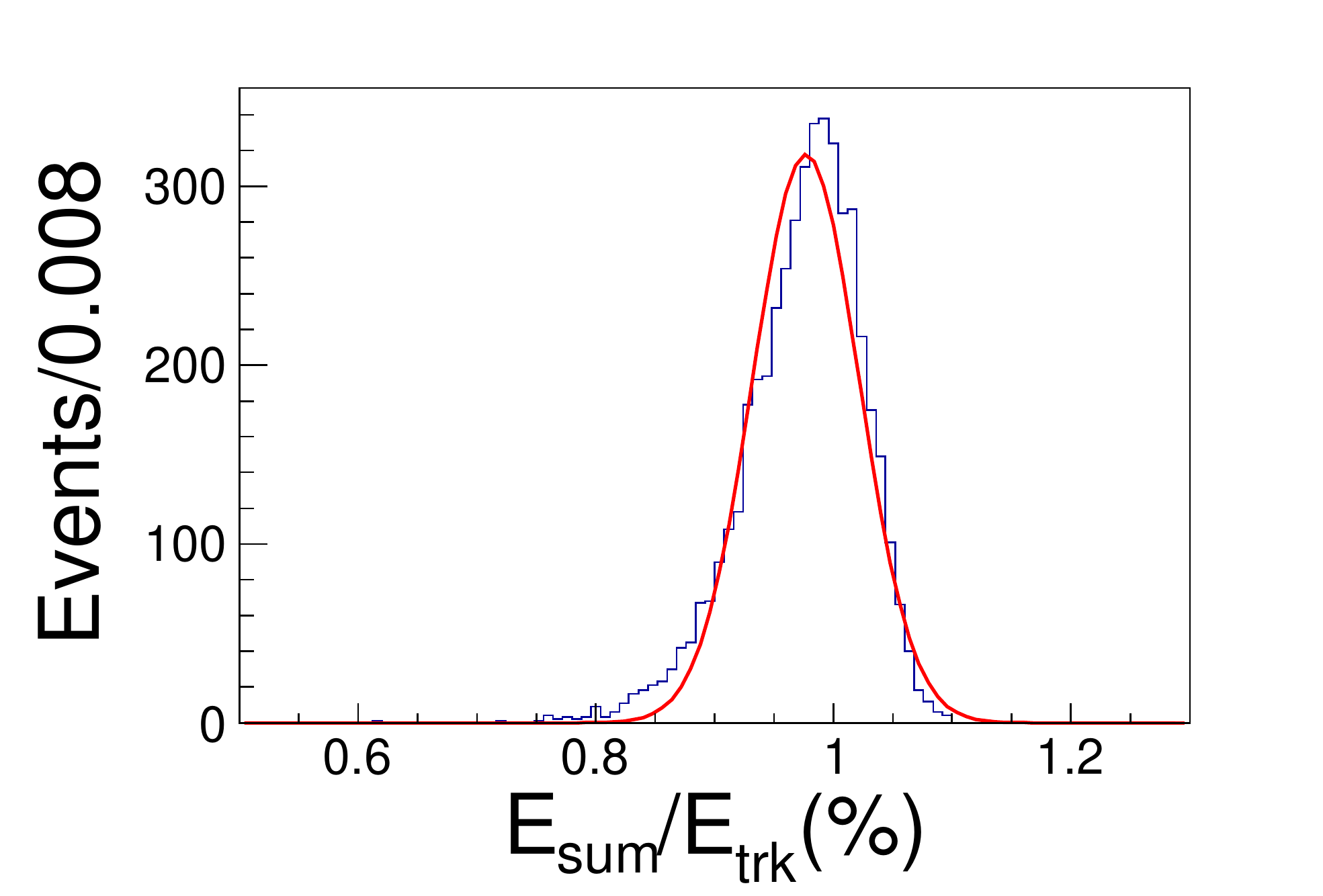}\put(-95,45){(${\rm E_{trk}}$ =140 GeV)}
\includegraphics[width=4.5cm]{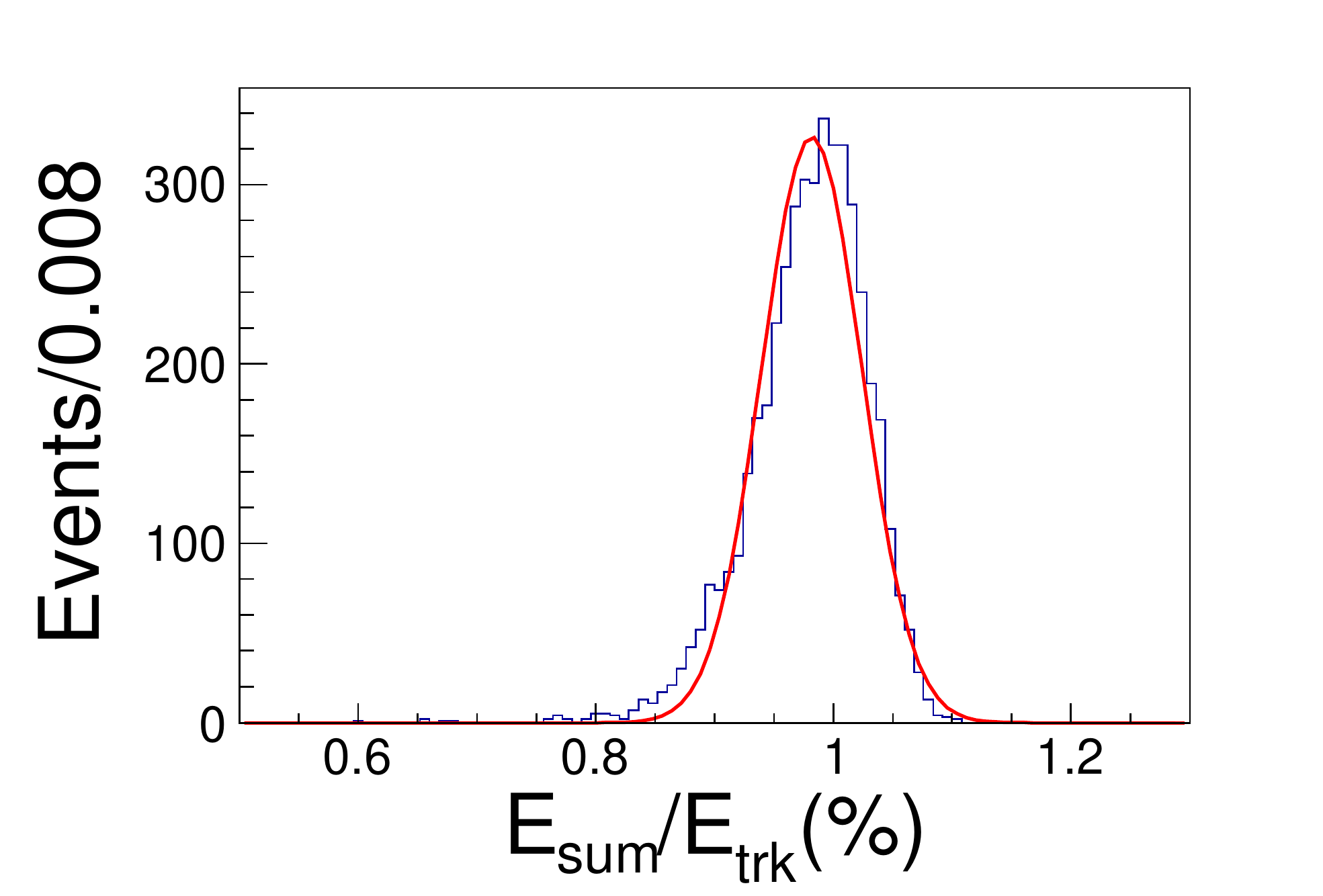}\put(-95,45){(${\rm E_{trk}}$ =160 GeV)}}
{\includegraphics[width=4.5cm]{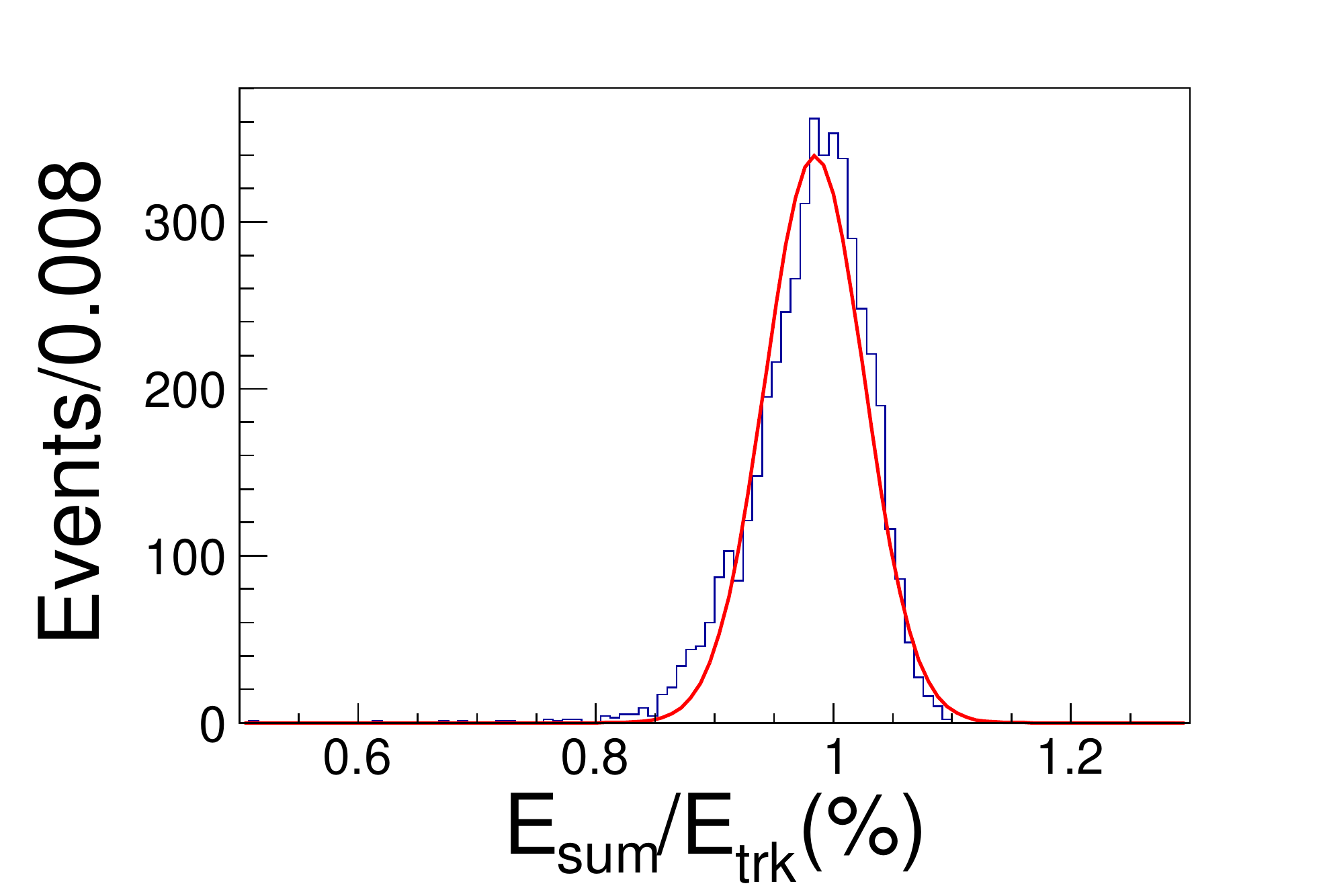}\put(-95,45){(${\rm E_{trk}}$ =180 GeV)}
\includegraphics[width=4.5cm]{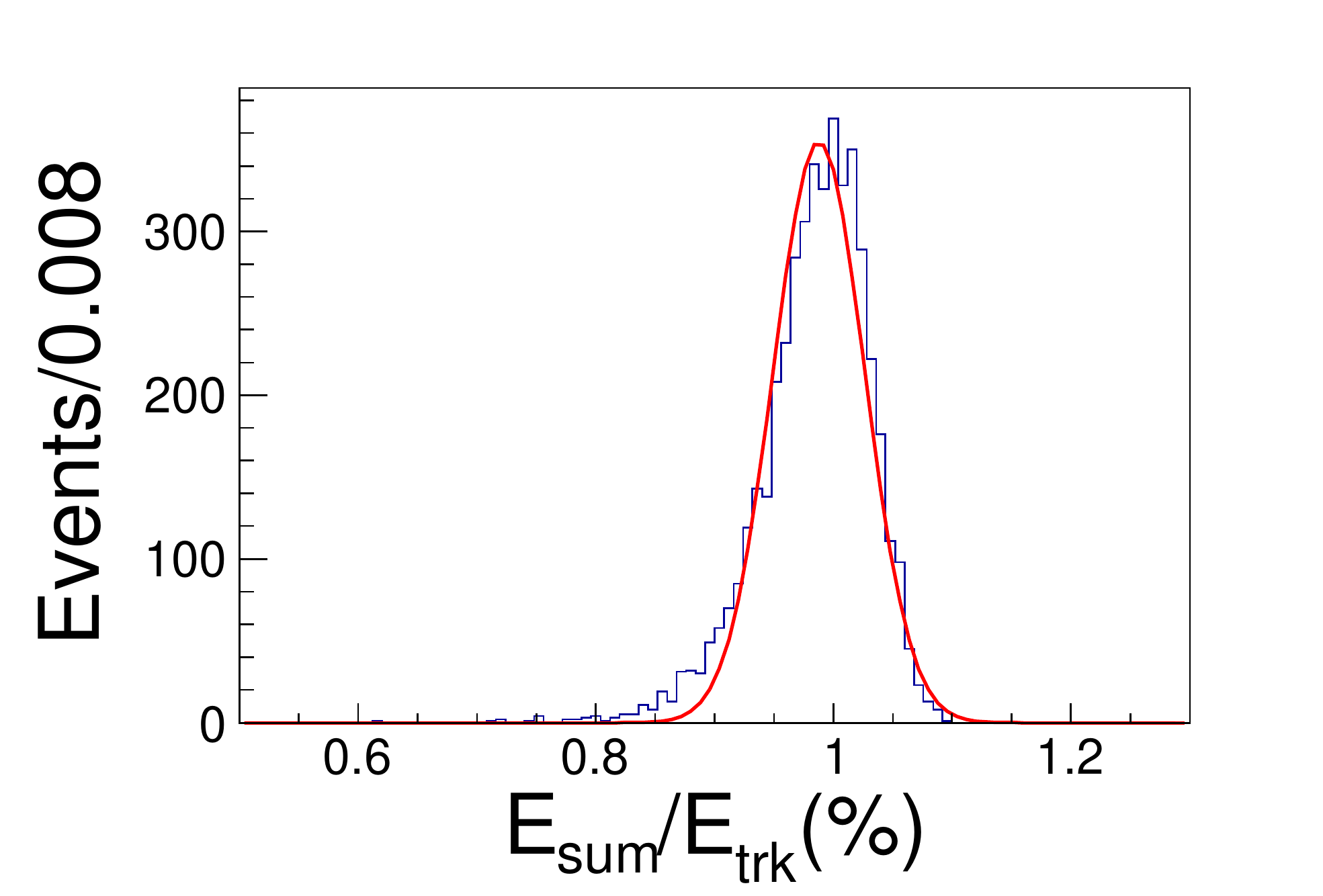}\put(-95,45){(${\rm E_{trk}}$ =200 GeV)}}
\figcaption{\label{cut_gap} (Color online) The blue histograms are the distributions of $\rm{E_{sum}/E_{trk}}$ for tracks shooting into non-gap regions with track energy vary from 20 to 200 GeV. The red curves are the fit lines with Gaussian functions, which give the mean and the width values of relative deposited energy.
}
\end{center}

\subsection{Direction reconstruction}
The center point coordinate $\rm{C(x_c,y_c,z_c)}$ of a track cluster is calculated by averaging over all the hits
in the global cluster,
\begin{equation}\label{ave_center}
\rm{C} = \rm{\frac{\Sigma_i C_i\cdot W_i}{\Sigma_i W_i}},
\end{equation}
where $\rm{C_i(x_i, y_i, z_i)}$ is the coordinate of each hit cell center determined by the detector structure and geometry, $\rm{W}_i$ is the weight function defined by
\begin{equation}\label{weight_func}
  \rm{W_i = max\{0, \mathscr{C} + \ln\frac{E_i}{E_{all}}\}}.
\end{equation}
Here $\rm{E}_i$ is the individual cell energy, and $\rm{E}_{all}$ is the sum of the energies of all the cluster cells,  $\mathscr{C}$ is a constant optimized to be $6$,  for minimizing the polar angle resolution (seeing in
Fig.~\ref{CONST_Optimize}), which means the low energy deposited hit cell (smaller than 0.25\% of
total deposited energy) has a weight factor 0.

\begin{center}
\includegraphics[width=4.5cm]{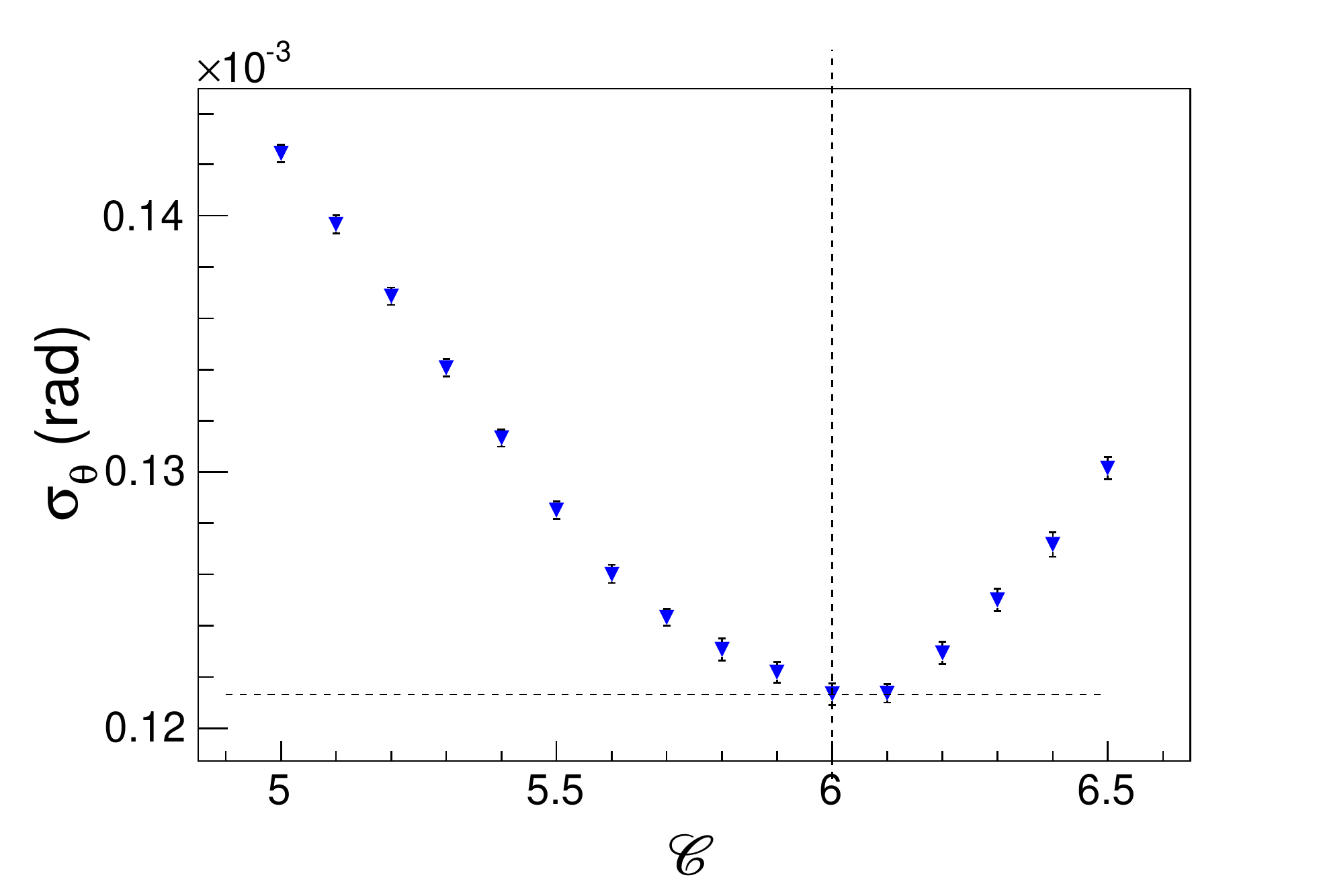}\put(-85,65){(a)}
\includegraphics[width=4.5cm]{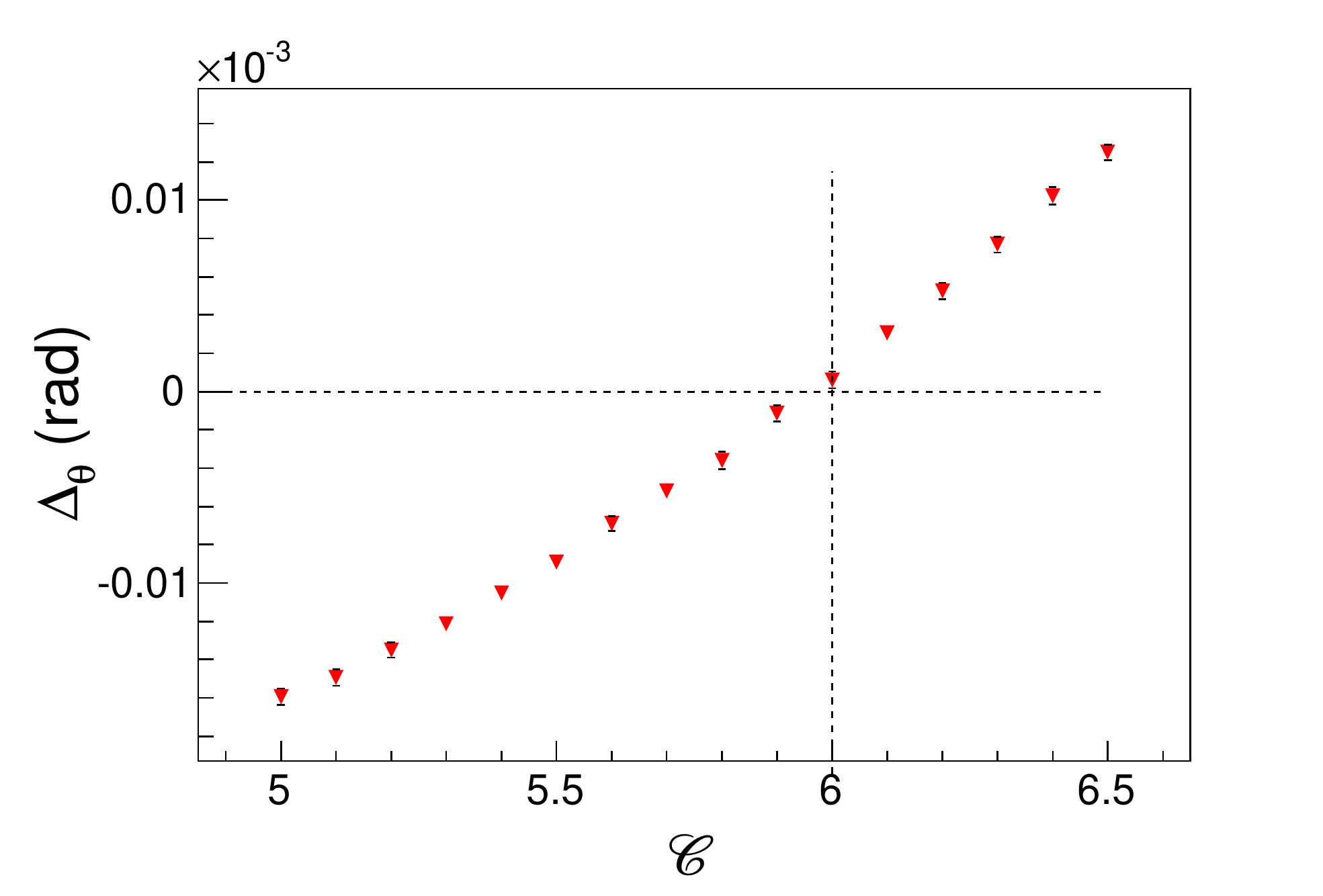}\put(-85,65){(b)}
\figcaption{\label{CONST_Optimize} (Color online) (a) The polar angle resolution, $\sigma_{\theta}$, and (b) the polar angle bias, $\Delta_{\theta}$ as a function of the constant, $\mathscr{C}$. }.
\end{center}


Any direction from the interaction point to each cluster center
is taken as the track strike direction.
The distributions of the difference between the reconstructed and the original
polar angles and azimuthal angles are shown
in Fig.~\ref{Pos_rec}, and fitted with Gaussian functions. It turns out
that the reconstruction bias of the polar angle and azimuthal angle are $\Delta_{\theta}=(6.04\pm4.42)\times10^{-4}~\rm{mrad}$ and
$\Delta_{\phi}(-1.75\pm2.08)\times10^{-2}~\rm{mrad}$, respectively; the resolution of polar angles and azimuthal angles are  $\sigma_{\theta}=(1.21\pm0.01)\cdot10^{-1}~\rm{mrad}$
and $\sigma_{\phi}=(5.86\pm0.02)~\rm{mrad}$.
\begin{center}
\includegraphics[width=4.5cm]{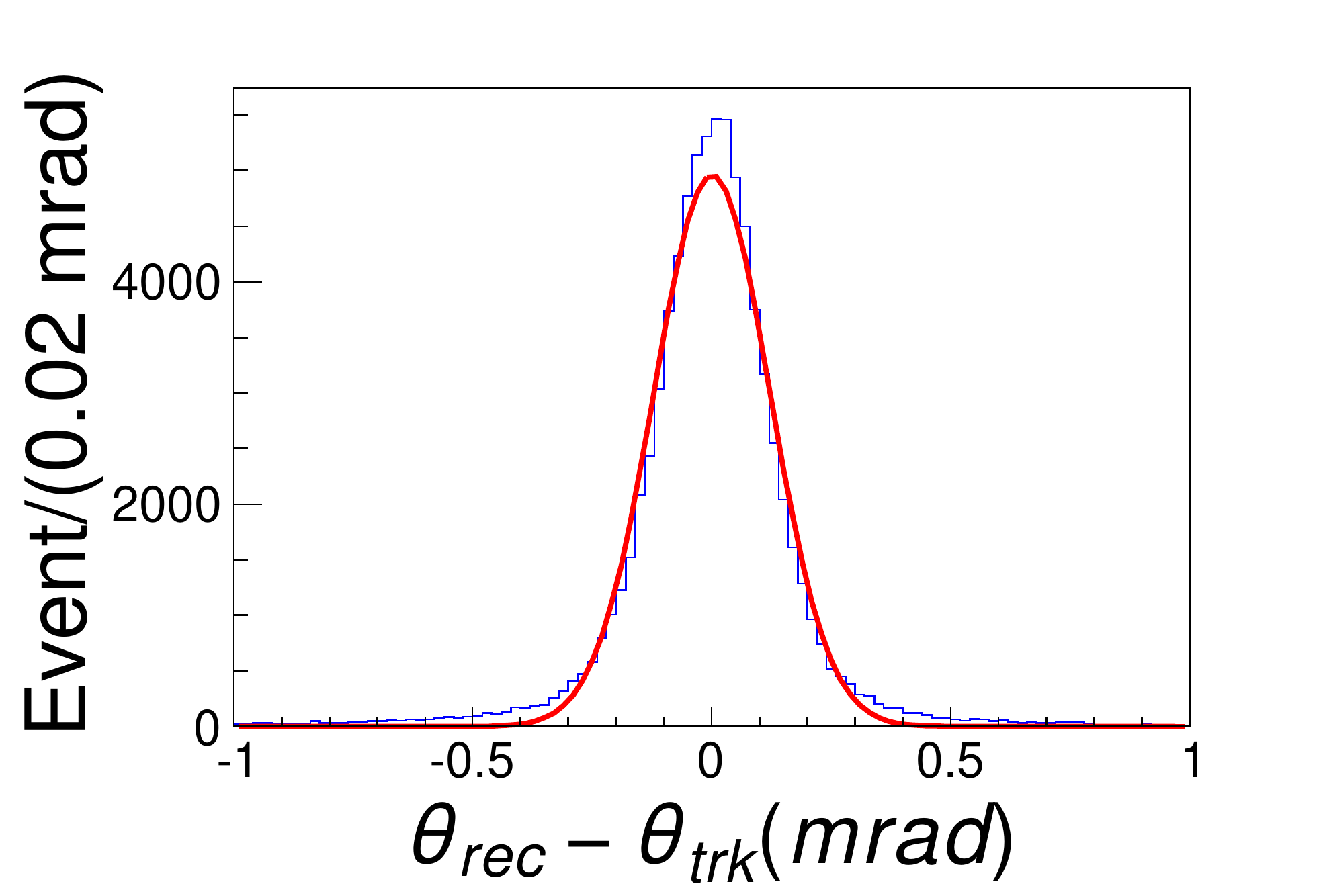}\put(-85,65){(a)}
\includegraphics[width=4.5cm]{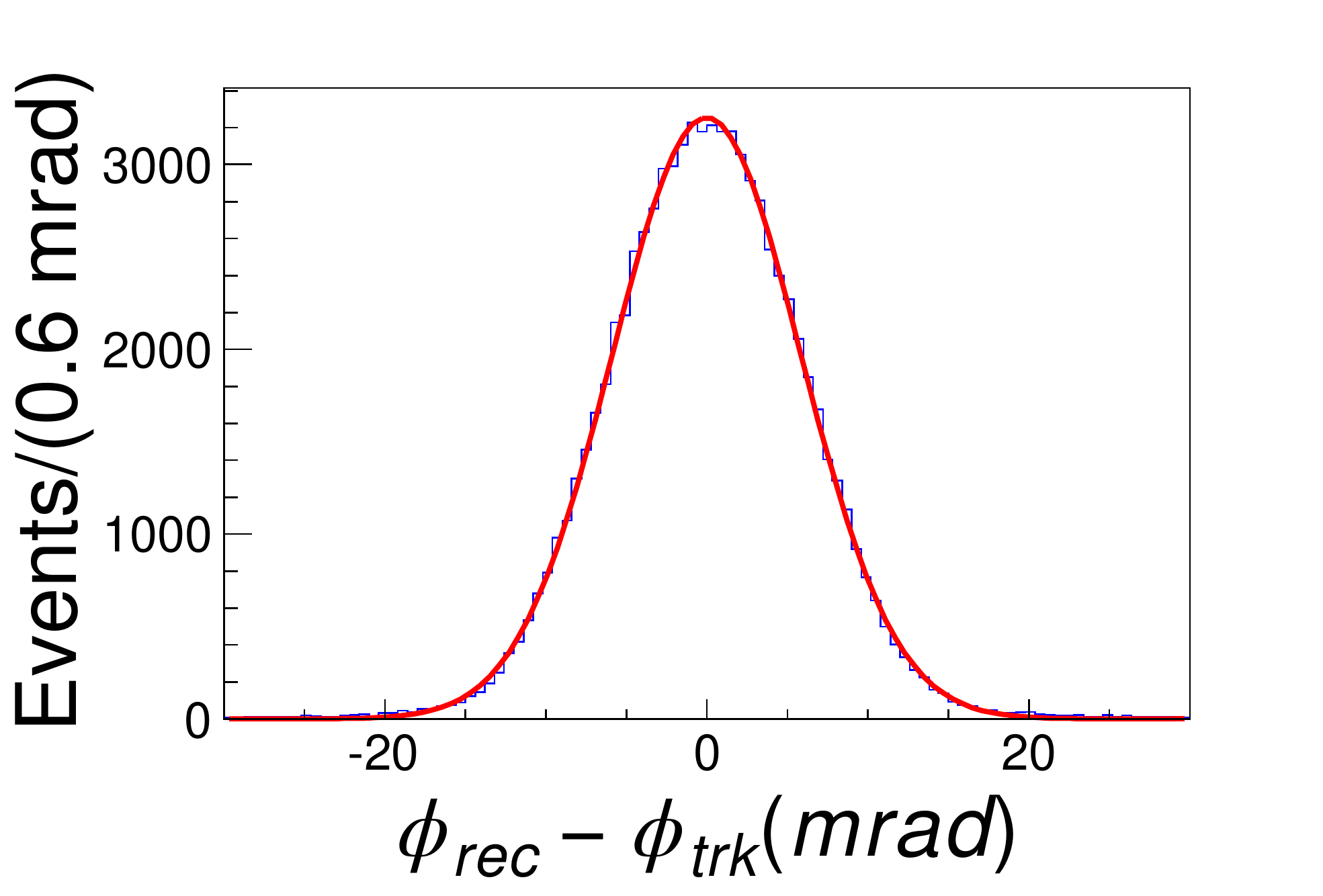}\put(-85,65){(b)}
\figcaption{\label{Pos_rec}  (Color online) The direction reconstruction result.
(a) The blue histogram is the distribution of $(\theta_{\rm{rec}}-\theta_{\rm{trk}})$, where the $\theta_{\rm{rec}}$ and
$\theta_{\rm{trk}}$ are the reconstructed and the original polar angles of the tracks, respectively.
 (b) The blue hisgogram is the distribution of $(\phi_{\rm{rec}}-\phi_{\rm{trk}})$,
where the $\phi_{\rm{rec}}$ and $\phi_{\rm{trk}}$ are the reconstructed and original azimuthal
angles of the tracks, respectively. The red lines are the fit results with the Gaussian functions. }.
\end{center}

\section {DNN based track reconstruction method}
\subsection{Energy reconstruction}
To deal with the tracks falling into the tile gap regions,
a neural network consisting $6$ fully connected layers are built for the energy reconstruction.
The architecture of the network is illustrated in Fig.~\ref{Dense_model}.

A neuron $i$ in a layer $l$ connects with a certain weight vector $\rm{w_{[i]}^{[l]}}$ to
all neurons in the following layers. The number of neurons per layer gradually decreases from
512 to 32 in the hidden layer 1 to hidden layer 5 with a factor of 2. Rectified Linear Units (ReLu) are chosen as activation functions throughout the system except to the input layer,
which is currently the most commonly used activation function~\cite{ref_DNN} and defined as
\begin{equation}
  {\rm ReLu}(x)=max(0, x).
\end{equation}
The energy
deposited values (in unit $\mathrm{GeV}$) in all sensor cells of one track cluster are
set as the input of our DNN model.
There are $30 \times 48 \times 60 = 92160$ cells in one module, thus the input of our network is a
($1 \times 92160$) matrix.
The output value of the model is taken as the energy prediction.
%
Before training, all weight matrix elements are initialized randomly with
uniform distribution of interval (-0.05, 0.05)
and bias values are initialled to 0.
\end{multicols}
\ruleup
\begin{center}
\includegraphics[width=13cm]{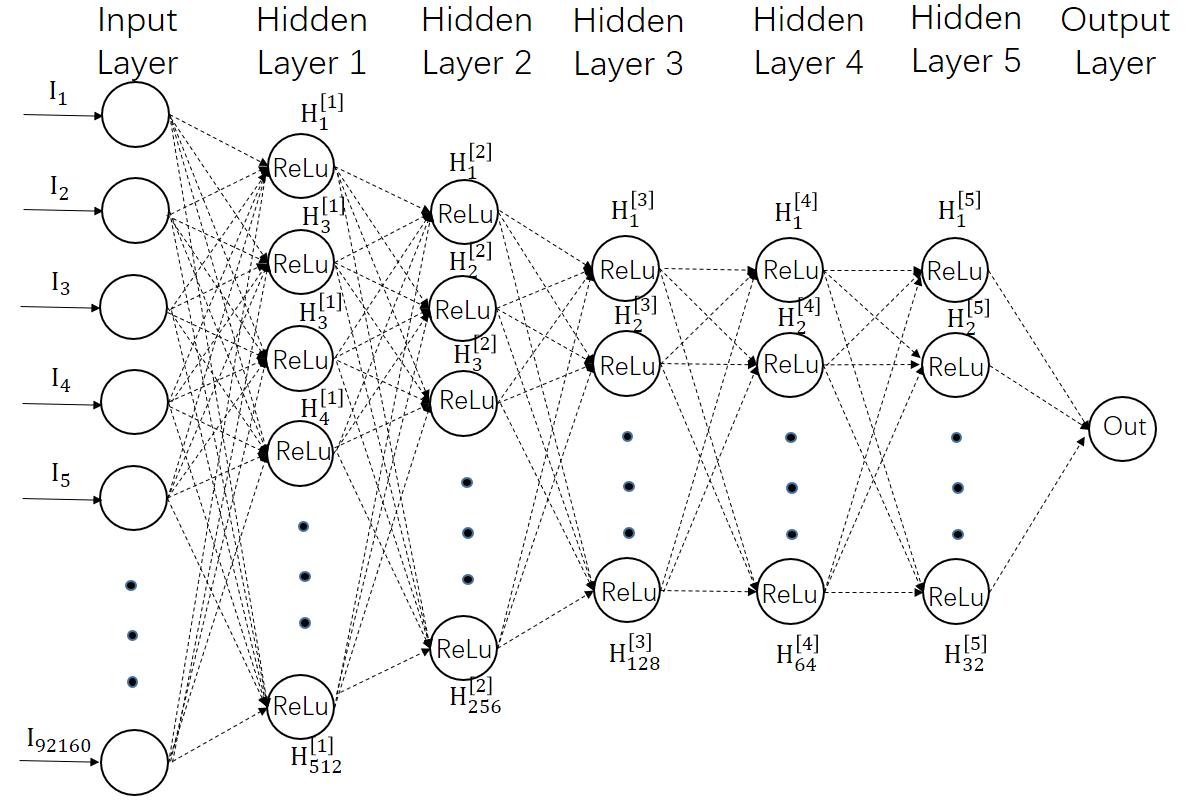}
\figcaption{\label{Dense_model}   Architecture of the DNN used for the energy reconstruction.
The input layer represents the 92,160 deposited energy values of all hit cells. It is fed to the
first hidden layer, which has 512 neurons with rectified linear unit activation functions. The
number of neurons per layer is gradually decreased from 512 to 256, 128, 64, 32, then to 1 in the output layer. The DNN is trained to obtain the true
energy in the output layer. The model contains 47,361,025 trainable parameters.}
\end{center}
\begin{multicols}{2}
To train the network we use mini-batch gradient descent to minimize the loss function. In
this approach, the gradient of the loss is computed using a subset of the training set, called a
mini-batch, which is chosen randomly during each minimization step. Therefore, it takes more than
one step to go through all the training set; the corresponding number of steps is called epoch.
We use a simple mean-square-error as the loss function of our model, which is defined as
\begin{equation}
  Loss = \frac{1}{m}\sum^{m}_{j=1}(y_j-\hat{y}_j)^2,
\end{equation}
where $m$, $y$, and $\hat{y}$ are the size of the mini-batch, the true of energy, and the
predict energy of event j, respectively.
Taking the $3\times10^5$ Monte Carlo events as the training set, the DNN model
is trained using the
{\sc Adam}~\cite{adam} optimizer.
The model is  trained for $30$ epochs in the rough training period with
learning rate 0.001 and another $20$ epochs with learning rate 0.0001 for
final training. During training, the size of mini-batch is set to 100.
The total number of neurons is 47,361,025 for the 6 layers DNN model. Most of the neurons
are placed in the first layer due to the large input dimension. The total number of neurons
will not change too much even if we add or delete some layers with small dimension. In fact, we tried
different neural network models from 1 layer to 20 layers. The experience tells that the more layers a DNN model has,
the faster to train but the model became more easily to be overfitting, and oppositely, the less layers the slower to train, thus 6 layers is a proper balance. Especially, the shallow net is very unstable, that indicates a simple weight method is difficult to be implemented reliably.

\begin{center}
\includegraphics[width=8cm]{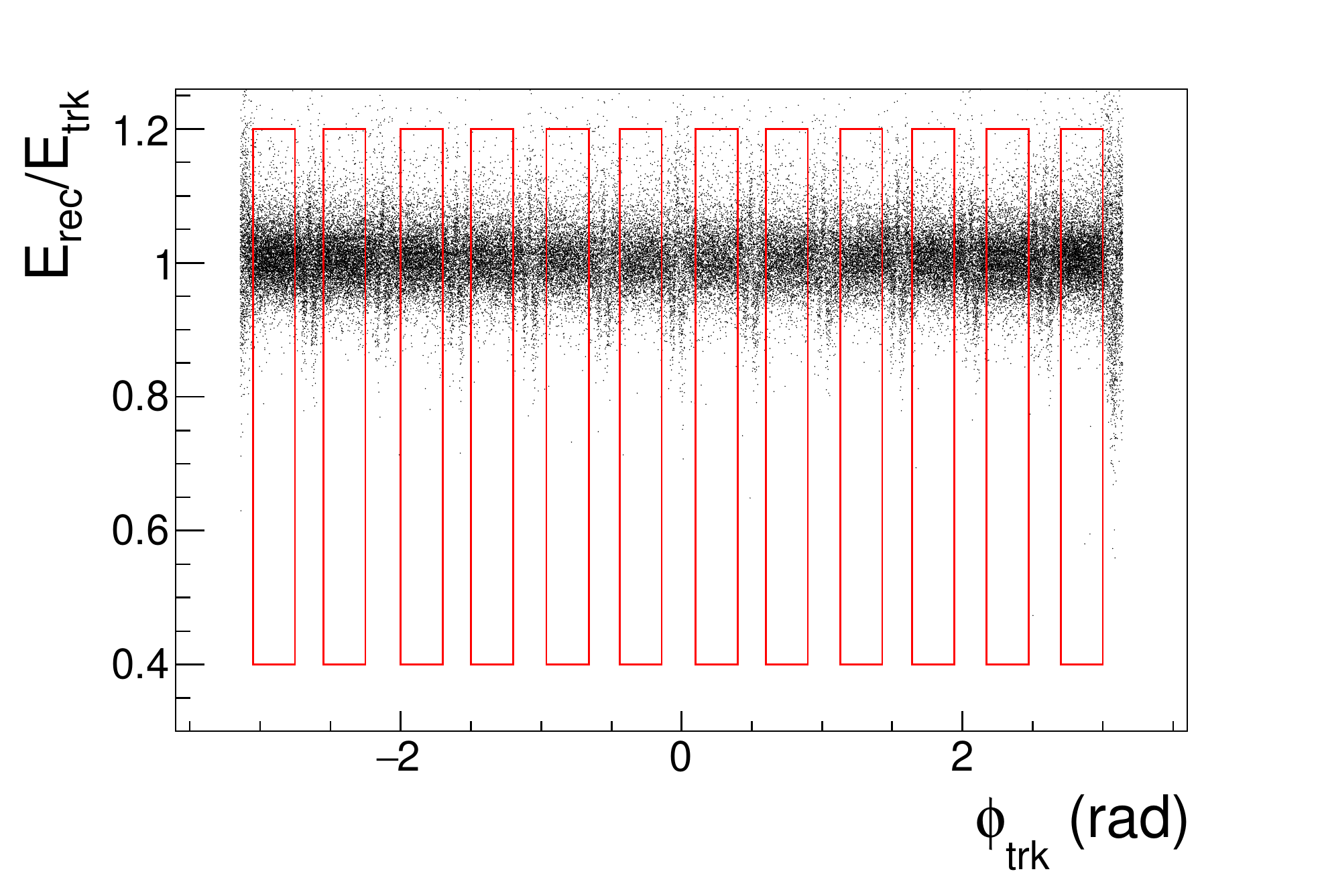}
\figcaption{\label{DNN_gap_eff} (Color online) The reconstructed energy by DNN method ($\rm{E_{rec}}$) divided by
the origin track energy ($\rm{E_{trk}}$) versus the incident track azimuthal
angle ($\mathrm{\phi_{trk}}$).
}
\end{center}

Different from the conventional method, we do not discard any events falling into tile gap regions
neither in training nor testing. The reconstructed energy divide the original
energy of incident track versus the azimuthal angle is shown in Fig.~\ref{DNN_gap_eff}.
The DNN reconstructed energy versus the origin track energy for different
energies are illustrated in Fig.~\ref{DNN_Erec}.

\begin{center}
{\includegraphics[width=4.5cm]{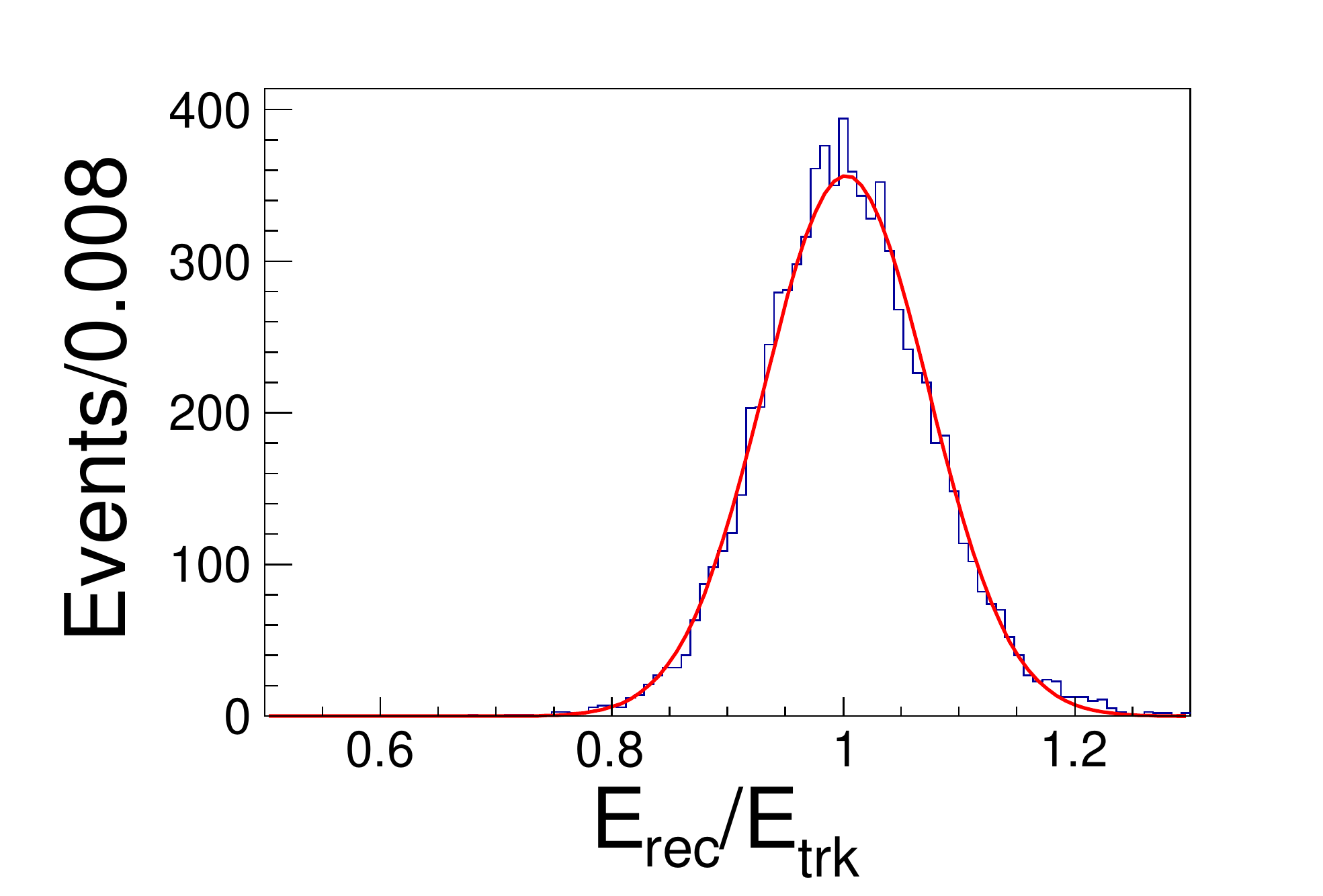}\put(-95,45){(${\rm E_{trk}}$ =20 GeV)}
\includegraphics[width=4.5cm]{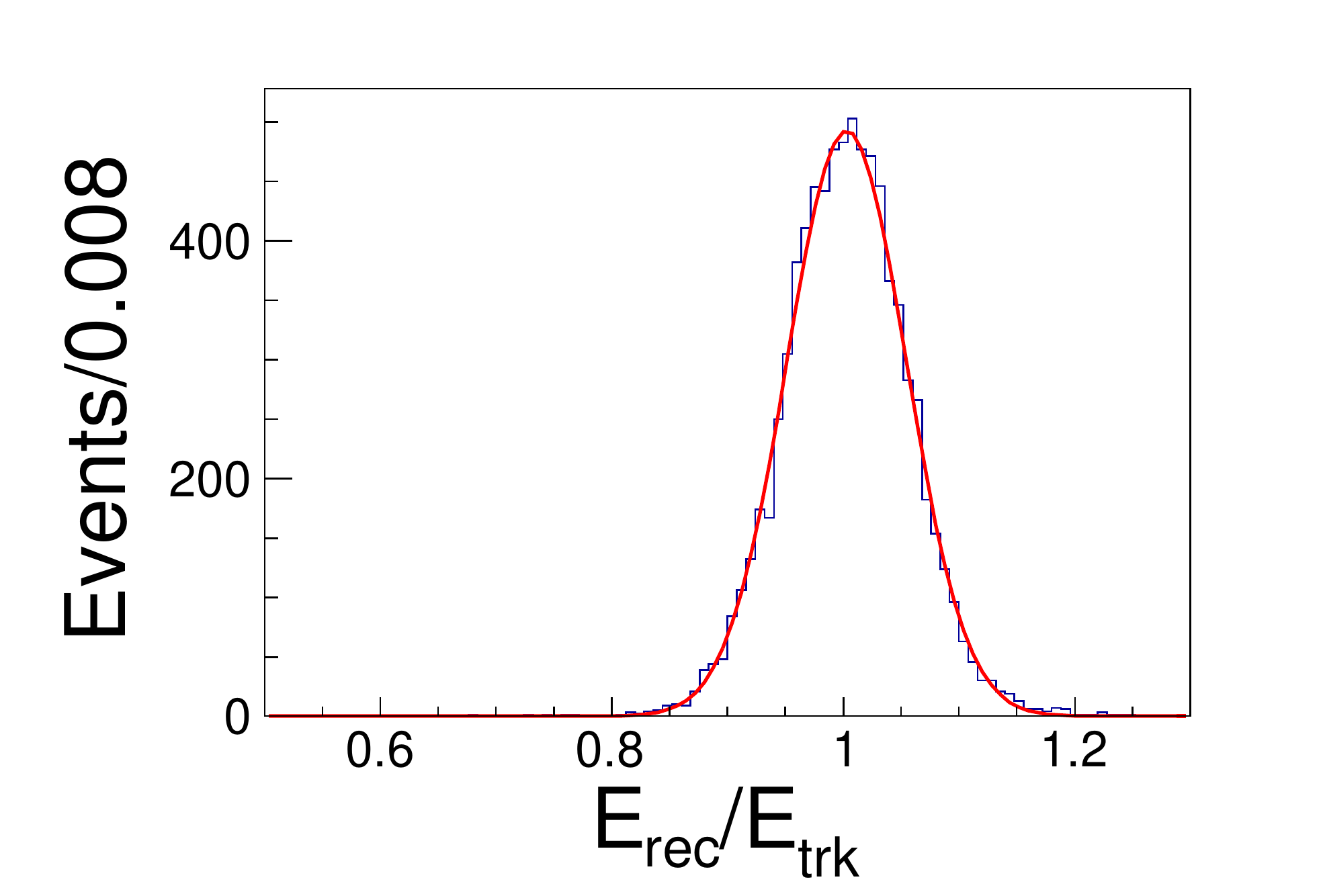}\put(-95,45){(${\rm E_{trk}}$ =40 GeV)}}
{\includegraphics[width=4.5cm]{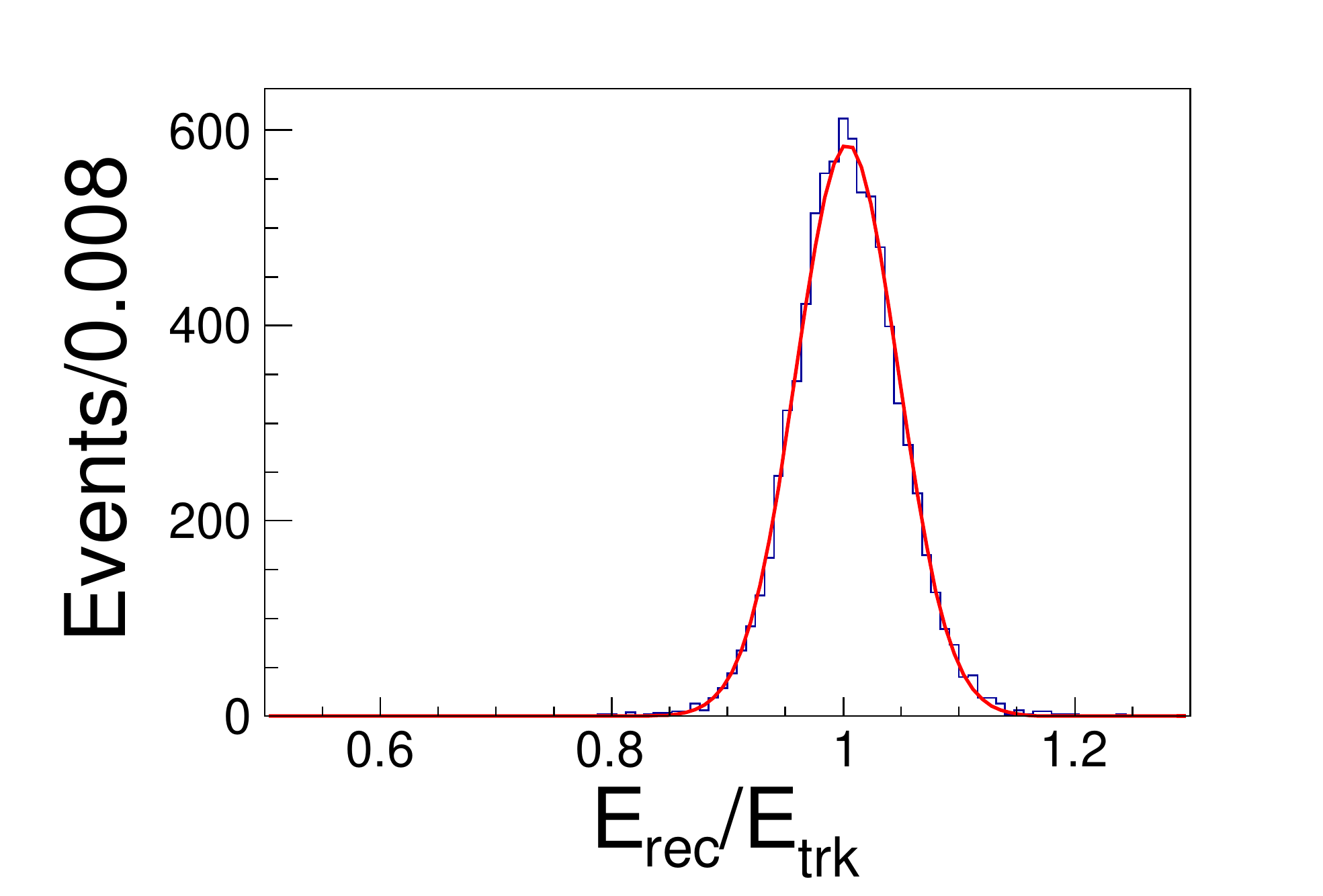}\put(-95,45){(${\rm E_{trk}}$ =60 GeV)}
\includegraphics[width=4.5cm]{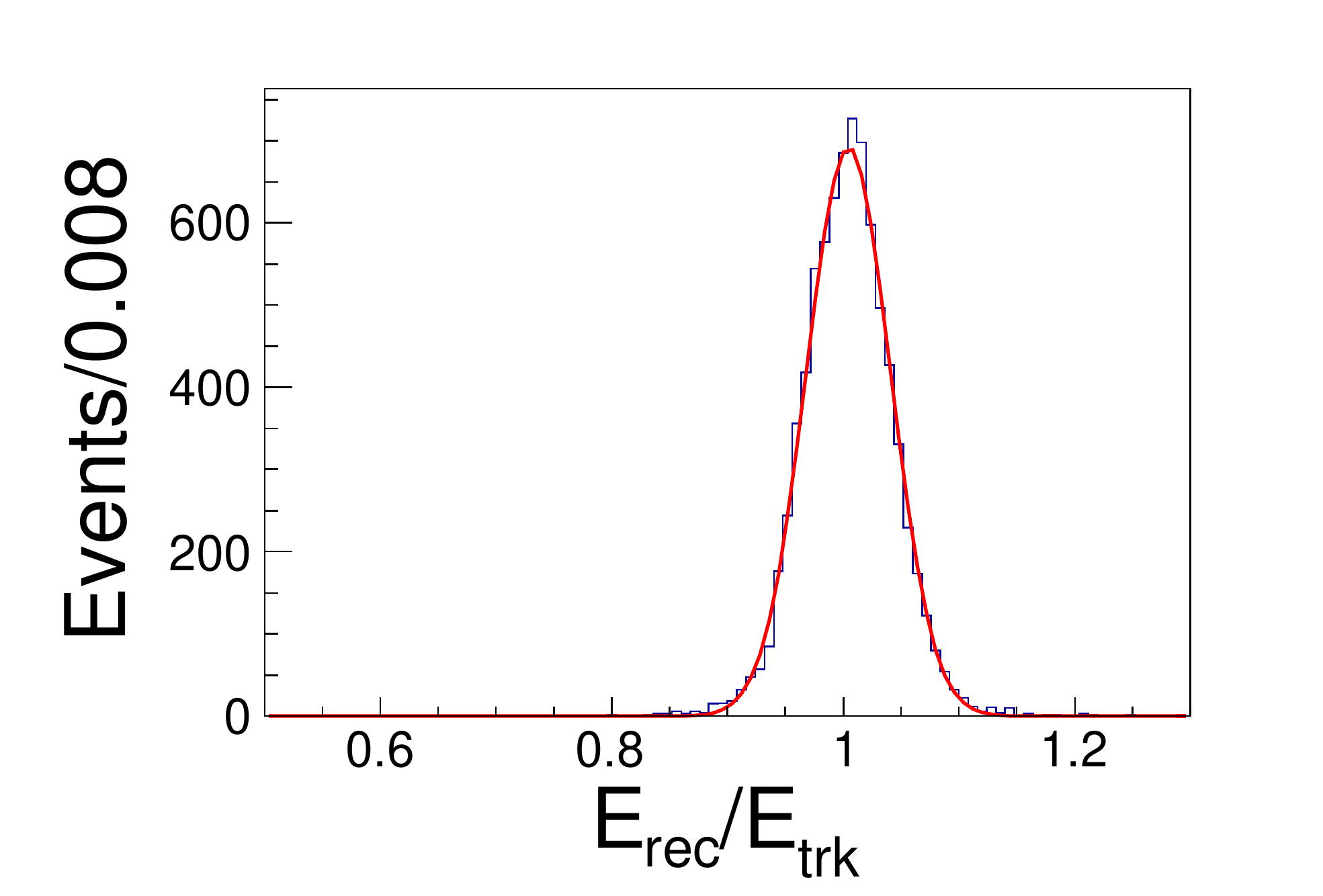}\put(-95,45){(${\rm E_{trk}}$ =80 GeV)}}
{\includegraphics[width=4.5cm]{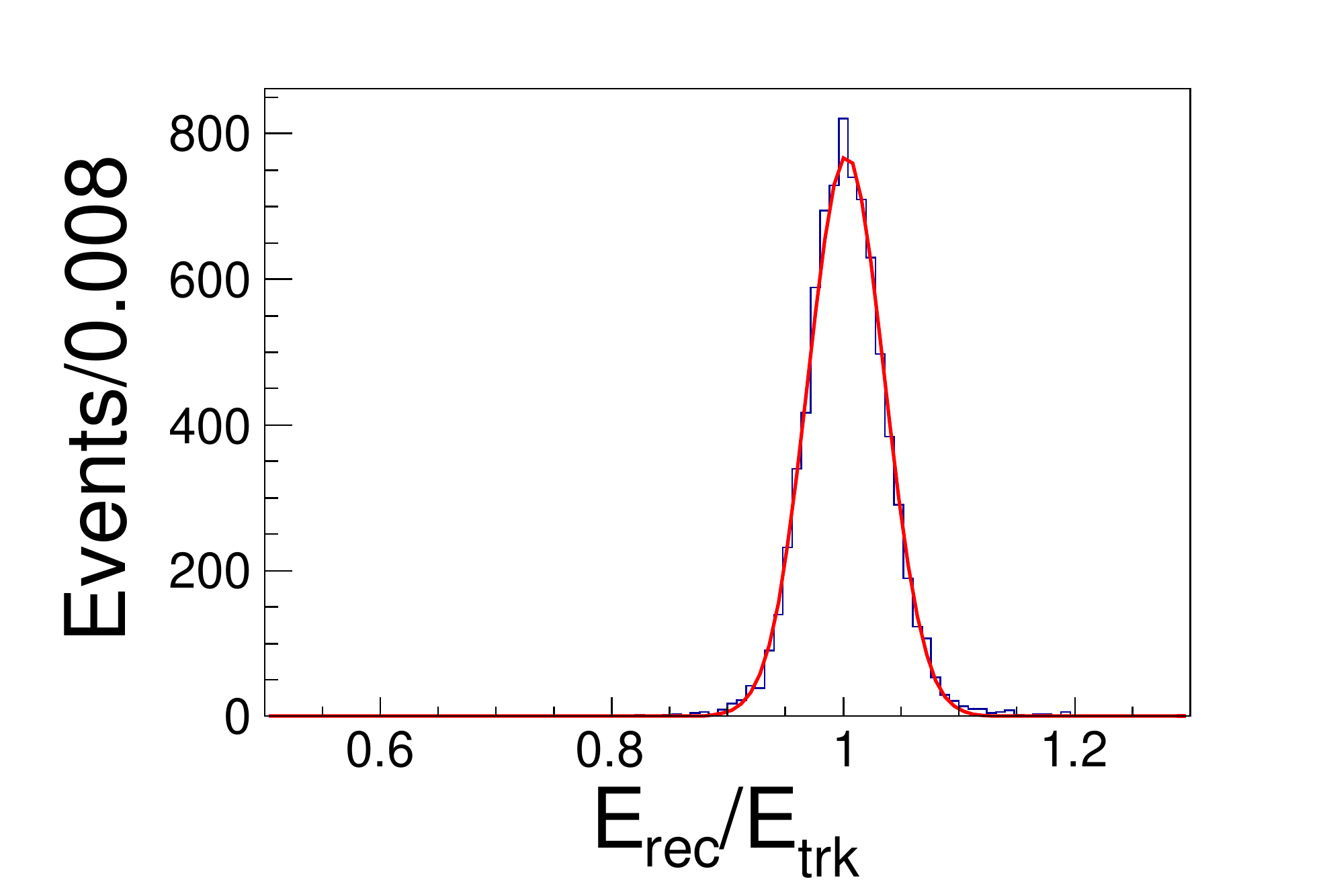}\put(-95,45){(${\rm E_{trk}}$ =100 GeV)}
\includegraphics[width=4.5cm]{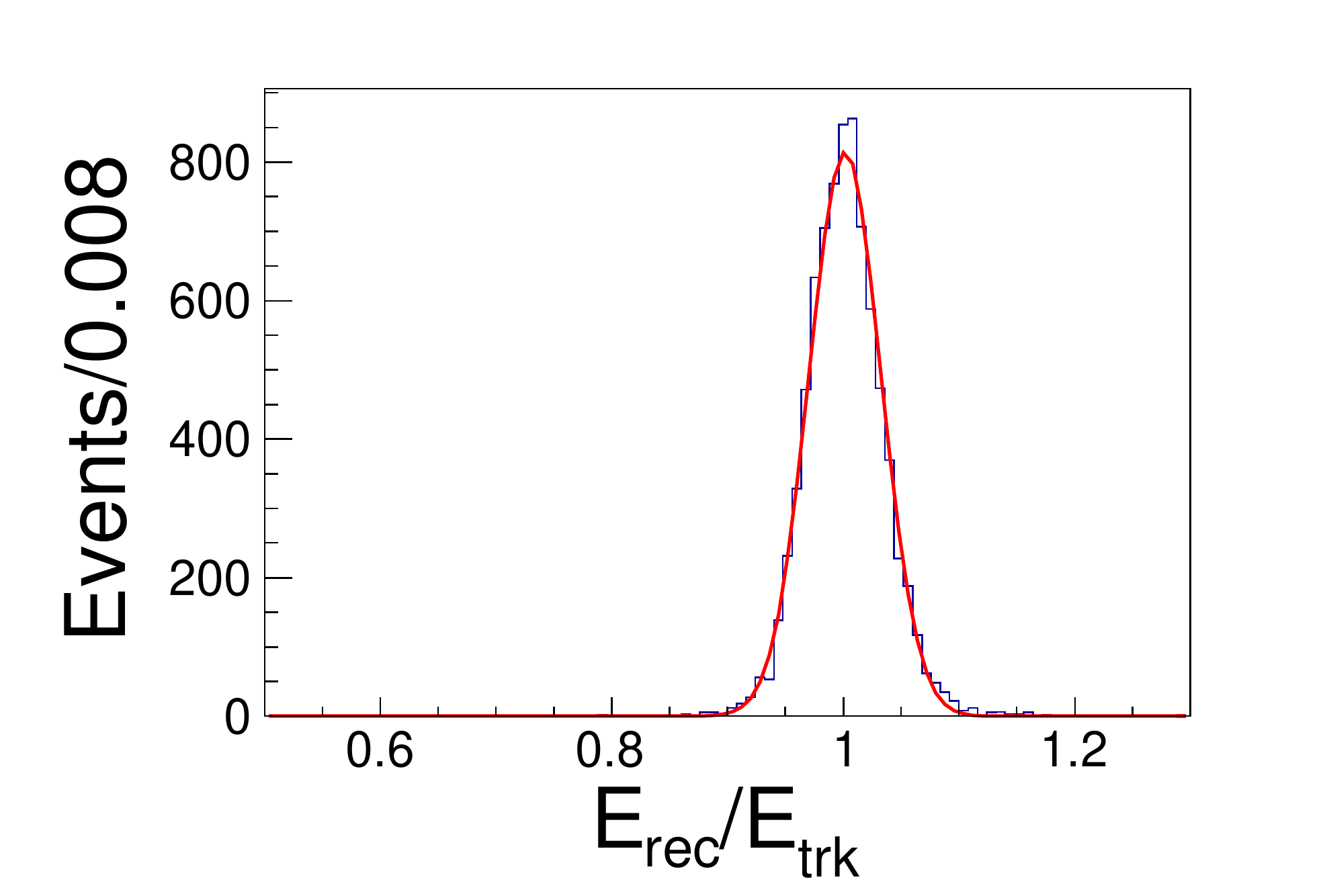}\put(-95,45){(${\rm E_{trk}}$ =120 GeV)}}
{\includegraphics[width=4.5cm]{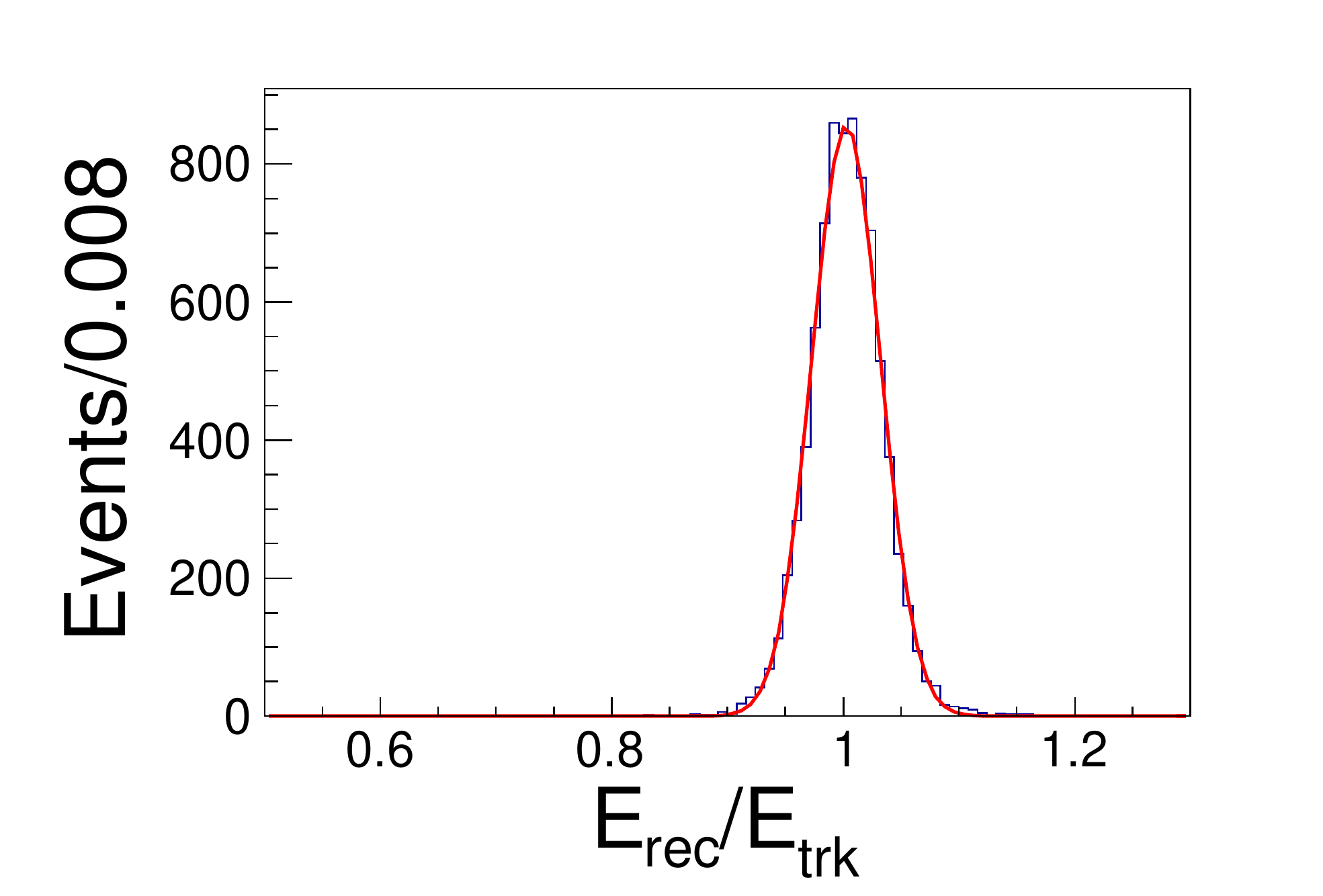}\put(-95,45){(${\rm E_{trk}}$ =140 GeV)}
\includegraphics[width=4.5cm]{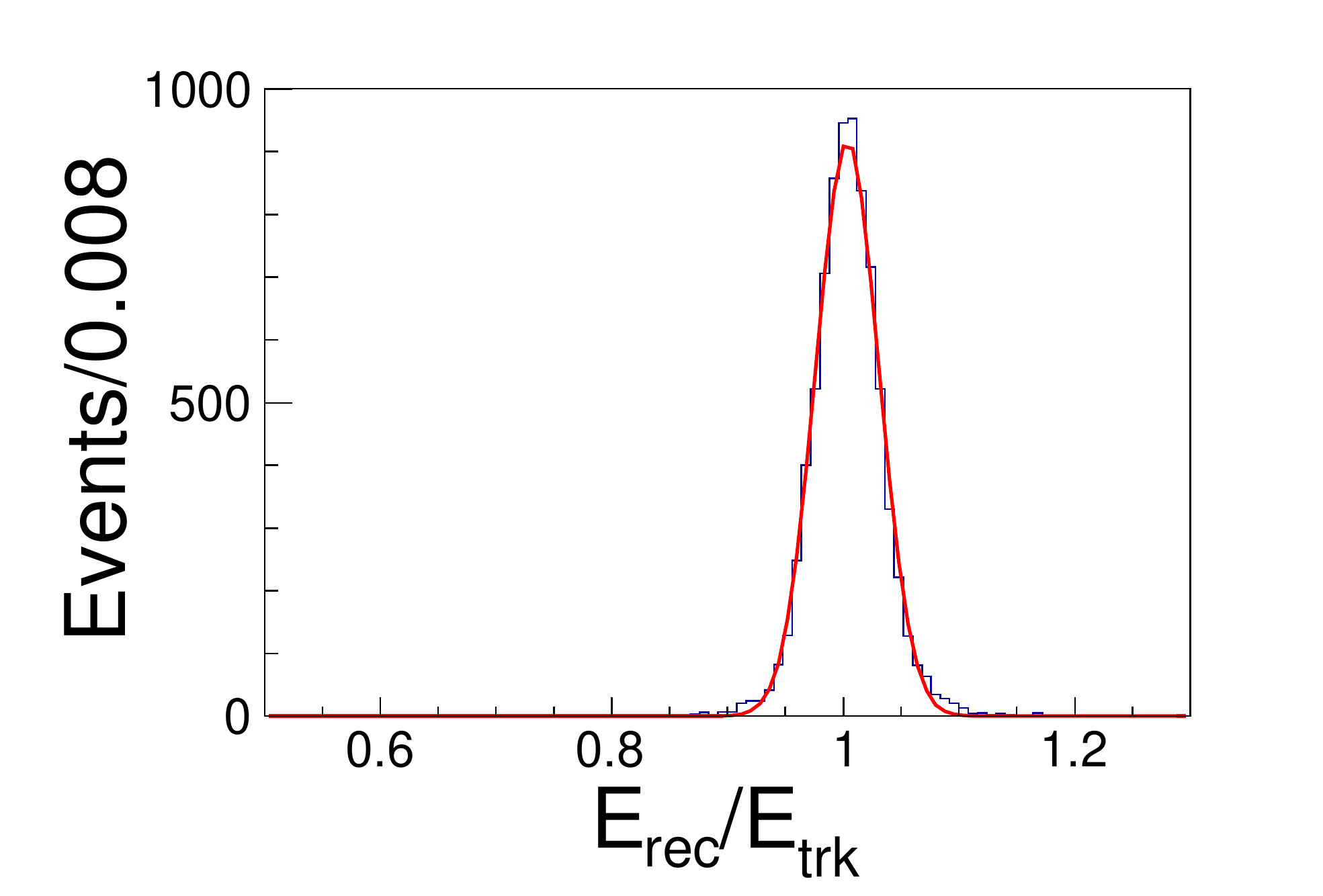}\put(-95,45){(${\rm E_{trk}}$ =160 GeV)}}
{\includegraphics[width=4.5cm]{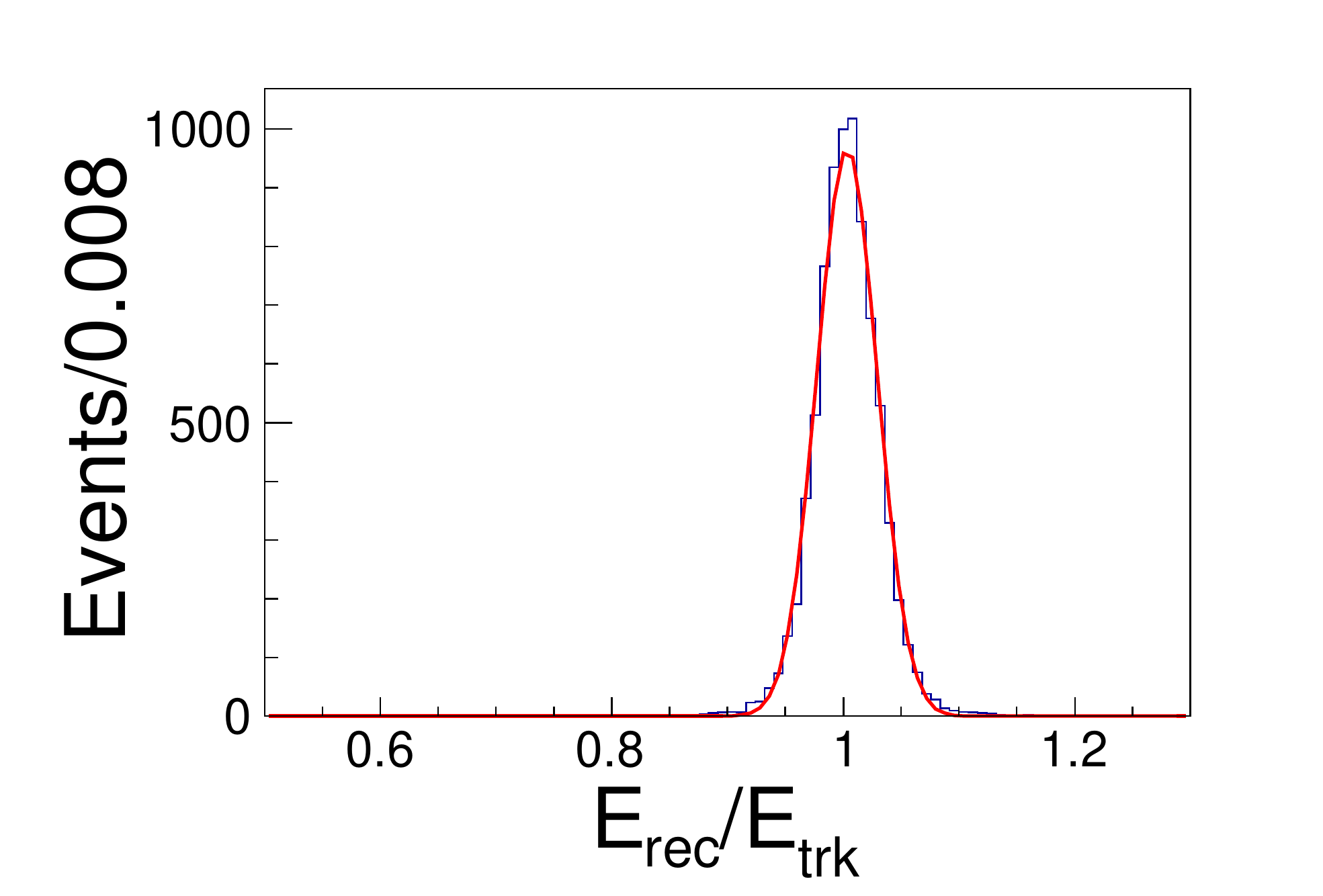}\put(-95,45){(${\rm E_{trk}}$ =180 GeV)}
\includegraphics[width=4.5cm]{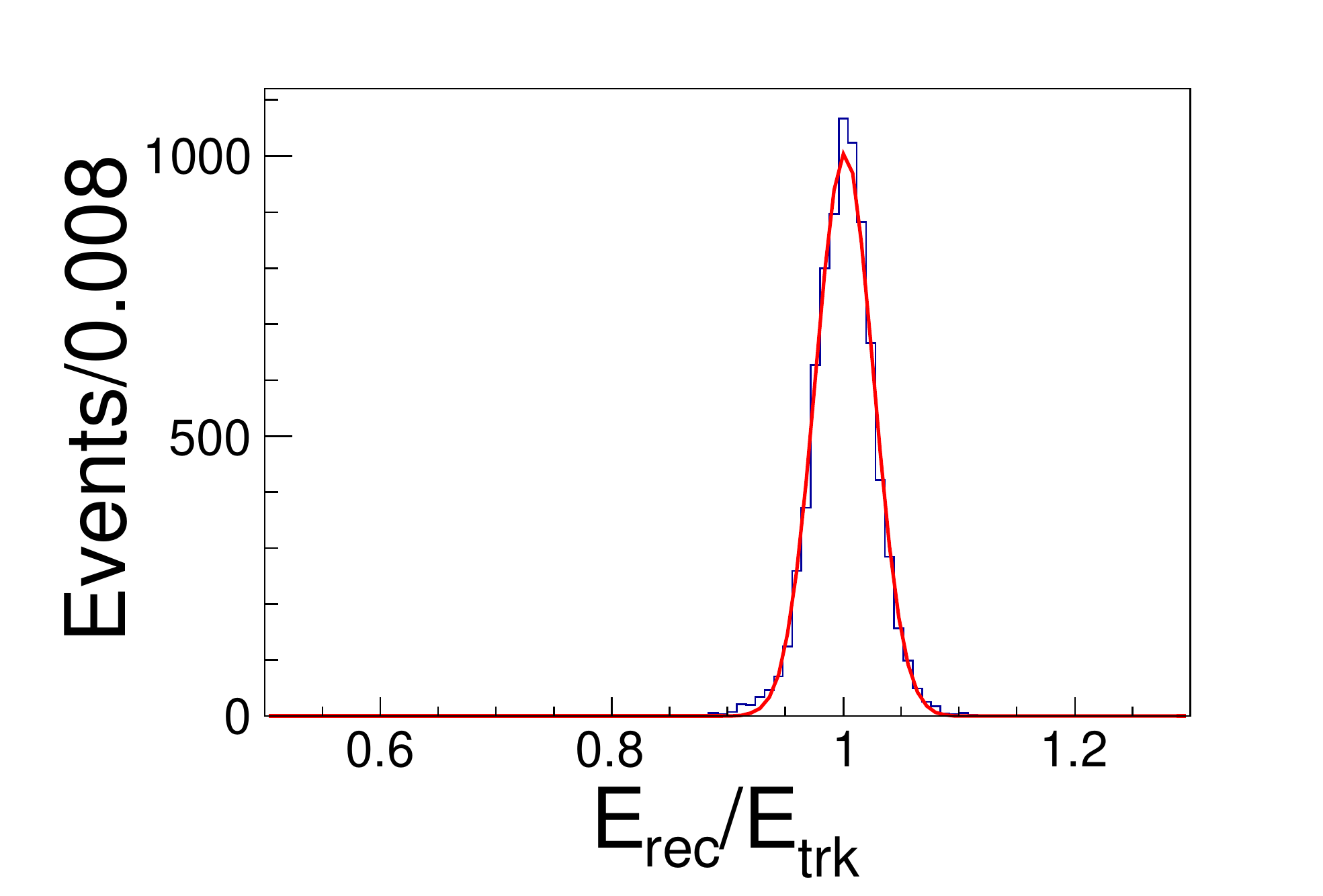}\put(-95,45){(${\rm E_{trk}}$ =200 GeV)}}
\figcaption{\label{DNN_Erec} (Color online) The distributions of $\rm{E_{rec}}/\rm{E_{trk}}$ from DNN method for track energy varying from 20 to 200 GeV.
The red lines are the fit results with gaussian functions.
}
\end{center}

\subsection{Direction reconstruction}
We reuse the same 6-layers DNN model architecture in Fig.~\ref{Dense_model}
to reconstruct the tracks incident direction ($\theta$, $\phi$).
To avoid the periodicity of azimuthal angle, we choose x-y coordinate
at z = 1000 mm as the output value of the model prediction (in unit mm).
In order to make the model easier to train, we construct two
models with same structure but training separately for predicting x and y coordinate
other than one model
with two dimension output to predict x and y simultaneously. Then we combine the
result of x prediction and y prediction
to calculate the polar angle and azimuthal angle.
Also  the energy deposited values (in units GeV) regard as a matrix
with shape $(1\times92160)$ are set
as the input of these DNN models and mean-square-error is chosen as the loss
function for these two models.
We find the normal initial strategy
will lead to vanishing gradient problem for these two models and
the transfer learning can deal with the problem well, which
the weight matrix elements of these two models are initialized with that of the
trained energy reconstruction model.
When training, the size of mini-batch is set to 100; the
learning rate decreases from $1.0\times10^{-2}$ to $1.0\times10^{-6}$.
These two models are trained  with the {\sc Adam} optimizer~\cite{adam} for about 100 epoches on the $3\times10^5$ Monte Carlo events till the loss function of the model does not decrease.
Using these two trained models, the reconstruction biases of polar angle and azimuthal angle can reach to $\Delta_{\theta} = (-5.65 \pm 0.84) \times 10^{-3}~\mathrm{mrad}$ and $\Delta_{\phi} = (-0.97 \pm 1.66) \times 10^{-2}~\mathrm{mrad}$, respectively;
the resolutions of polar angle and azimuthal angle  can reach to $\sigma_{\theta} = (2.34\,\pm\,0.01) \times10^{-1}~\mathrm{mrad}$ and $\sigma_{\phi} = (4.61\,\pm\, 0.02)~\mathrm{mrad}$, respectively, as shown in Fig~\ref{DNN_Pos_rec}.
\begin{center}
\includegraphics[width=4.5cm]{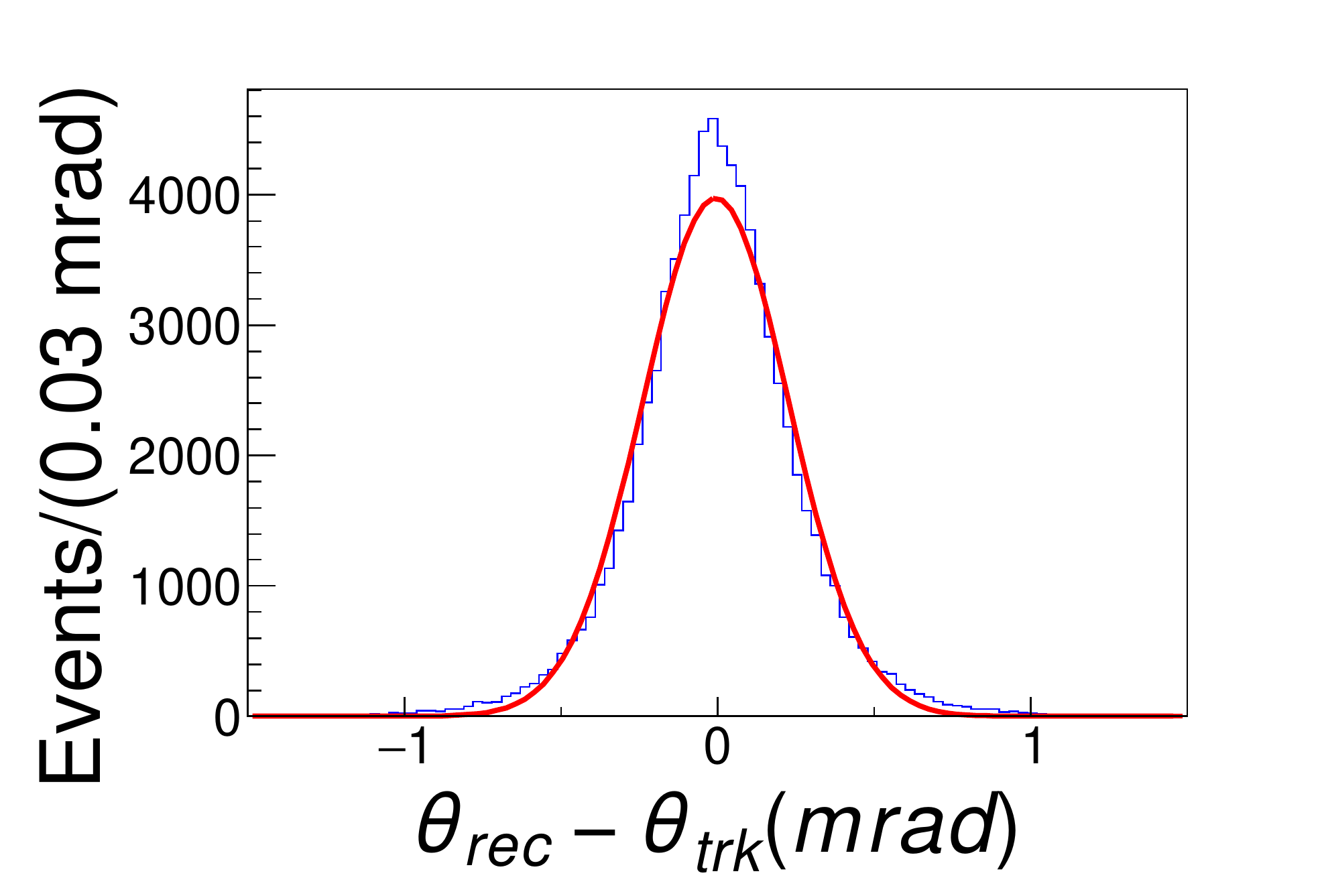}\put(-85,65){(a)}
\includegraphics[width=4.5cm]{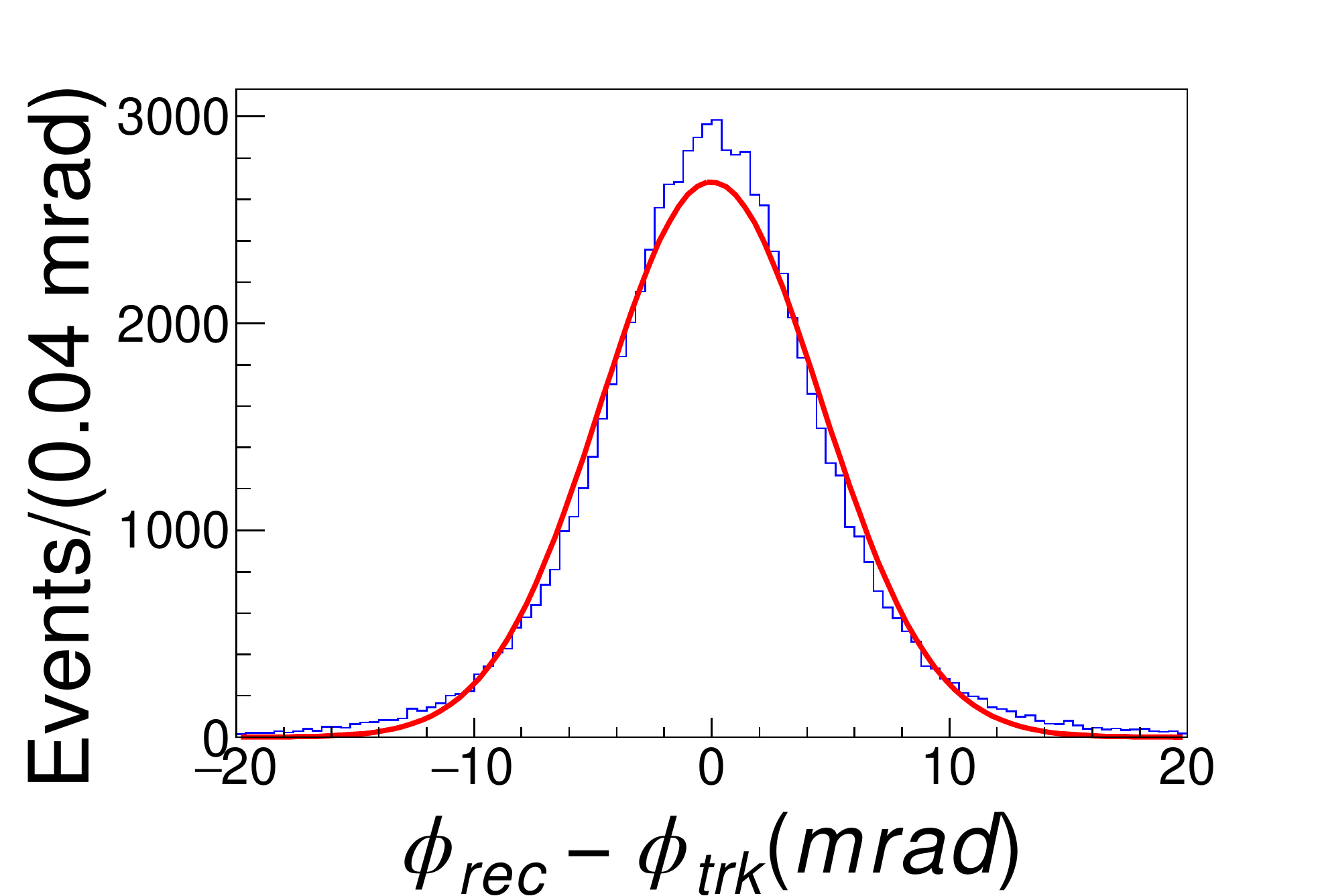}\put(-85,65){(b)}
\figcaption{\label{DNN_Pos_rec} (Color online) The direction reconstruction results.
(a) The blue histogram is the distribution of $(\theta_{\rm{rec}}-\theta_{\rm{trk}})$, where the $\theta_{\rm{rec}}$ and
$\theta_{\rm{trk}}$ are the reconstructed and the original polar angles of the tracks, respectively.
 (b) The blue histogram is the distribution of $(\phi_{\rm{rec}}-\phi_{\rm{trk}})$,
where the $\phi_{\rm{rec}}$ and $\phi_{\rm{trk}}$ are the reconstructed and original azimuthal
angles of the tracks, respectively.  The red lines are the fit results with the Gaussian functions. }
\end{center}
\end{multicols}
\begin{center}
\includegraphics[width=17cm]{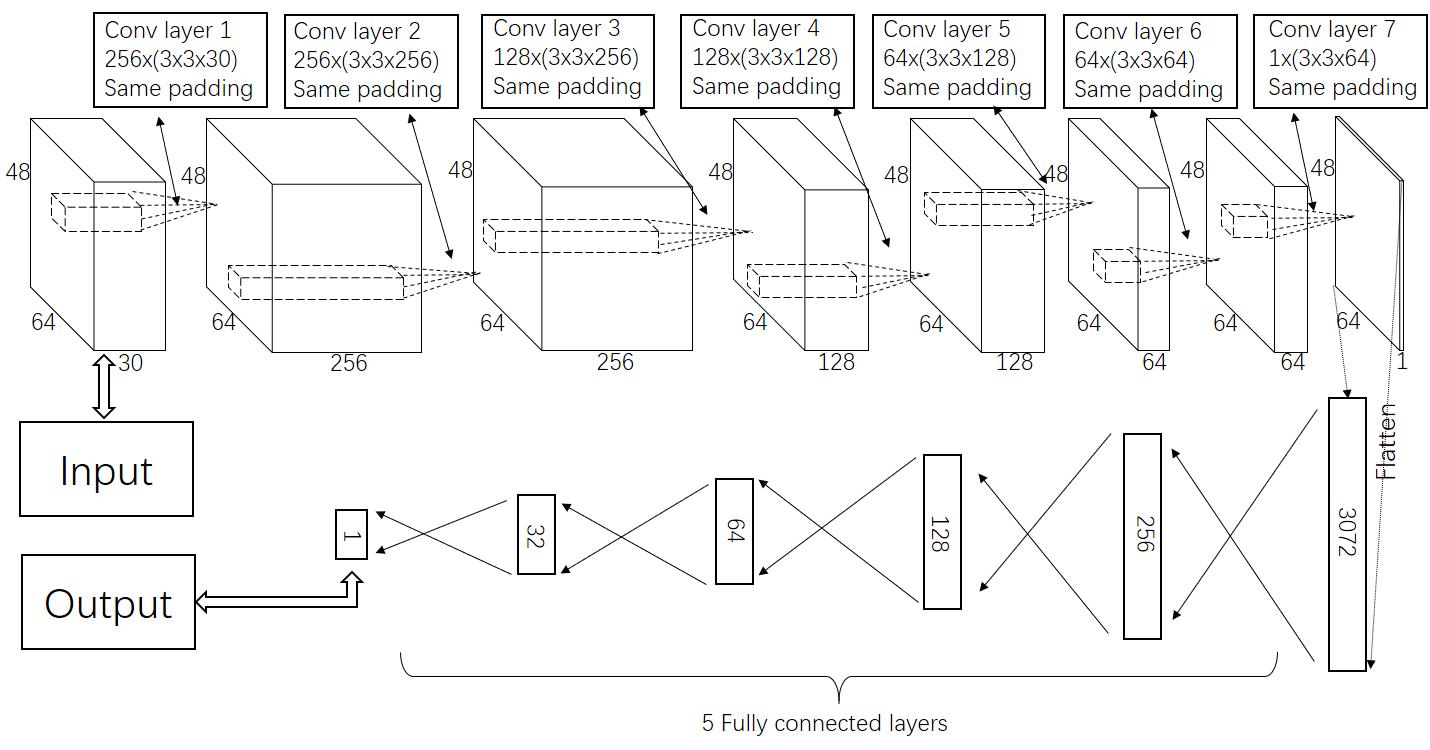}
\figcaption{\label{arch-cnn} Architecture of the deep neural network for the
polar angle reconstruction. The input layer represents the 92,160
(in shape ($30\times64\times48$)) deposited energy values of all hit cells.
It is followed with 7 convolutional layers. Each convolutional layer uses the same padding method
to keep the shape of output matrix  the same as the input matrix. The numbers of convolutional
kernels for each convolutional layers are 256, 256, 128, 128, 64, 64 and 1.
A flatten layer, which flatten the input matrix to one dimensional output matrix,
is following the last convolutional layer.
Following the flatten layer is 5 fully connected layers with a final output layer
consisting one unit, which is taken as the polar angle prediction. The model contains
2,043,330 trainable parameters.}
\end{center}

\begin{multicols}{2}
Comparing to the conventional method, we can get a better azimuthal resolution,
but worse polar angle resolution. To get a more precision polar angle resolution,
we construct a new DNN model which consists of 7
two-dimension convolutional layers, and 5 fully connected layers.
In all of the convolutional layers, the shape of the convolutional kernels are set to $N_c\times(3\times3)$, where $N_c$ denotes the number of channels for the input of each layer.
The energy deposited values (in unit GeV) are set as the input of this model with
a shape of $(30\times64\times48)$,
which is directly from the geometry structure of the luminometer.
Here 30 is taken as the number of channels for the model input.
ReLu activation function is used for all neurons of fully connected layers.
The architecture of the neural network is illustrated in Fig.~\ref{arch-cnn}.

The mean-square-error is taken as the loss function. The size of mini-batch is also set to 100
when training. All elements in weight matrix and convolutional kernels are initialized randomly
with uniform distribution of interval (-0.05, 0.05) and the bias values are initialled to 0.
Also taking the $3\times10^5$ Monte Carlo events as the training set,
the DNN model is trained sufficiently using the {\sc Adam}~\cite{adam} optimizer.
The trained model can predict the polar angle in
a bias of $\Delta_{\theta}=(4.22\pm0.35)\times10^{-3}~\mathrm{mrad}$ and a resolution of
$\sigma_{\theta}=(9.82\pm0.03)\times10^{-2}~\mathrm{mrad}$. The distribution of difference
between the reconstructed polar angle and the original polar angle is
illustrated in Fig~\ref{CNN_Theta_rec}.
\begin{center}
\includegraphics[width=6cm]{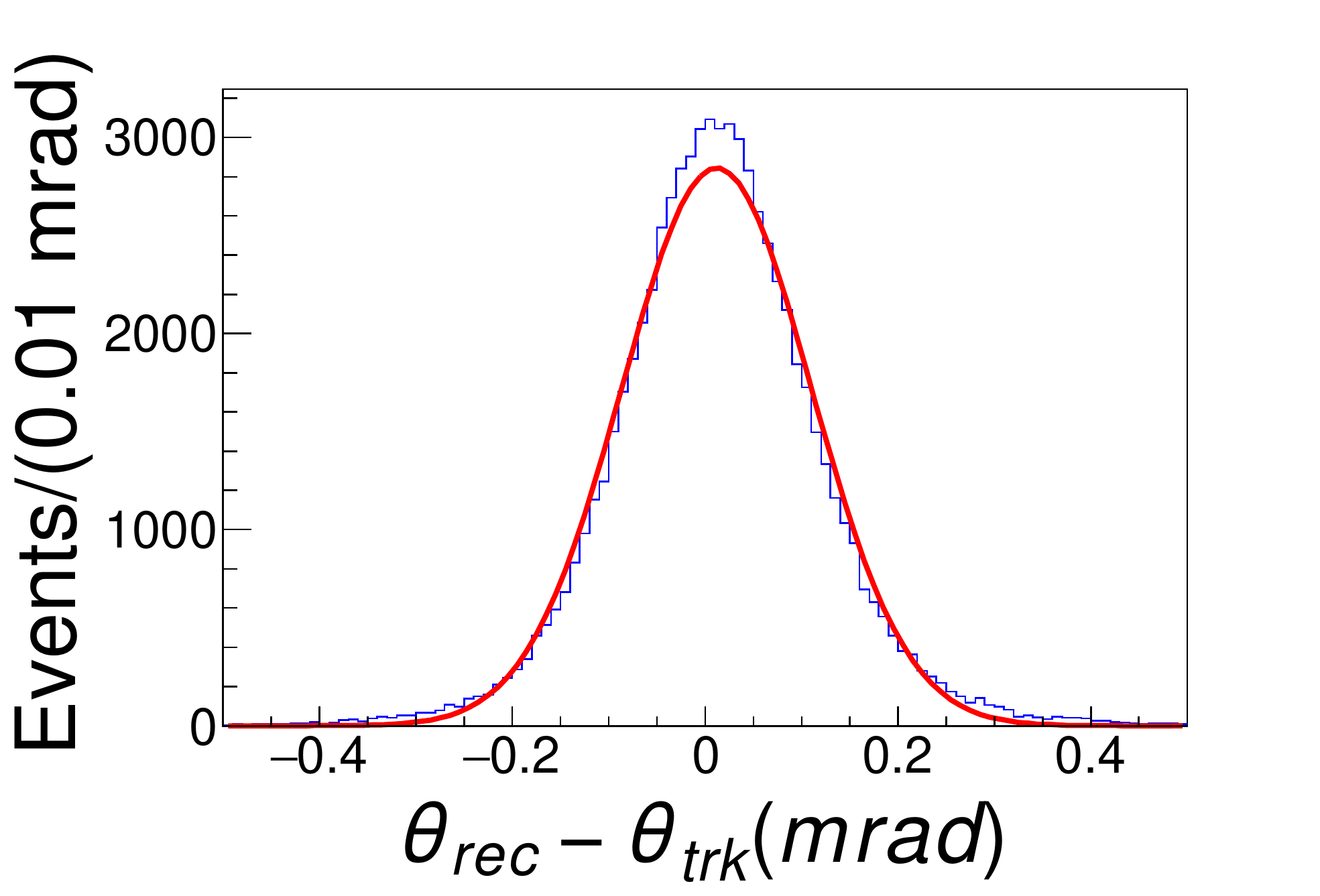}
\figcaption{\label{CNN_Theta_rec} (Color online) The blue histogram is the distribution of $(\theta_{\rm{rec}}-\theta_{\rm{trk}})$, where the $\theta_{\rm{rec}}$ and
$\theta_{\rm{trk}}$ are the reconstructed polar angle obtained from CNN model and the original polar angles of the tracks, respectively. The red line is the fit result with the Gaussian function.}
\end{center}

{\sc Keras}~\cite{keras} (v2.2.2) with the TensorFlow~\cite{tensorflow}
backend (v1.10.0) is used to construct all the models above. Four NVIDIA Tesla V100-SXM2 graphics cards are used for the training and testing.
\section{Comparison of two methods}
\subsection{Energy reconstruction performance}
Comparing Fig.~\ref{gap_eff}
and Fig.~\ref{DNN_gap_eff}, it is obvious that the DNN method can deal with the tracks falling in the tile
gap regions well. No obvious energy leakage is observed as in the conventional method.
Those results are
directly shown in Fig.~\ref{DNN_Erec}, where the $\rm{E_{rec}/E_{trk}}$
shapes from the DNN method are
obvious more symmetric and sharper compared with the conventional method shown in Fig.~\ref{cut_gap}.
The energy resolution distribution is fitted with following formula provided in Ref.~\cite{fabjan},
which is for the energy resolution of a sampling calorimeter defined as
\begin{equation}\label{res_func}
\frac{\sigma_{E}}{E}=\sqrt{\frac{a^2}{E}+b^2},
\end{equation}
where the first term corresponds to stochastic shower processes and the second corresponds to energy leakage.
In Fig.~\ref{res_compare}, we illustrate the energy
resolution of the two methods mentioned above, and fitting results with this formula.
Comparing to the conventional method, the DNN method gives better energy resolution
at all energy points.

\begin{center}
\includegraphics[width=8cm]{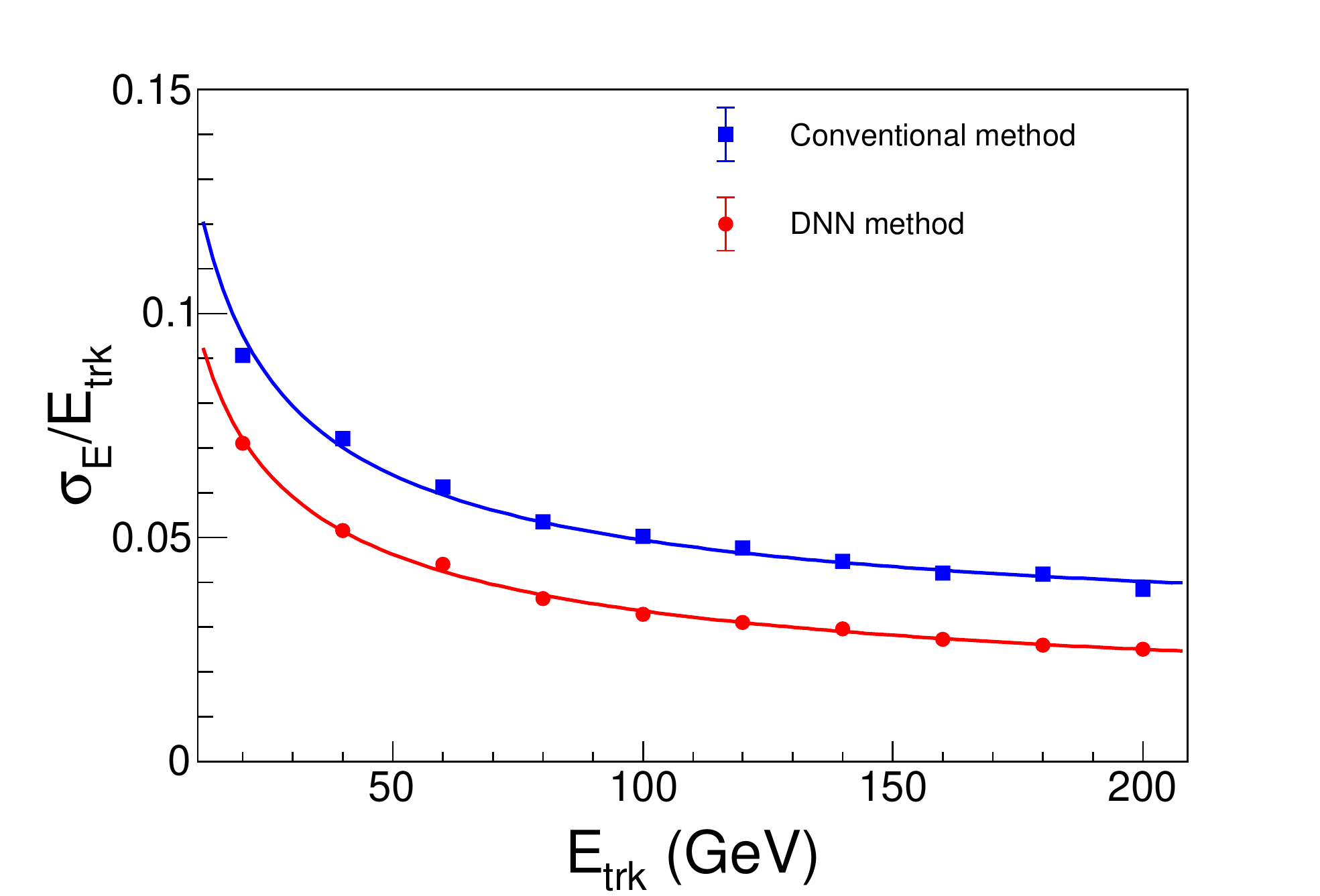}
\figcaption{\label{res_compare} (Color online) The comparison of energy resolution between two energy reconstruction methods. (tracks which shoots into tile gap are excluded only for conventional method.) The red (blue) line is the fit result with Eq.~(\ref{res_func}). For the red line, the value of parameter a is determined to 0.34, and b is $0.012$. For the blue line, a is 0.40, and b is 0.030. The error bars, caused by the Monte Carlo statistics, are too small to be seen at this scale.}
\end{center}

\subsection{Direction reconstruction performance}
In Table~\ref{direction_result}, we list the polar angle and azimuthal
angle reconstruction result of the conventional method and the DNN method.
Obviously, the DNN solutions can give smaller resolutions,
both for polar angle and azimuthal angle reconstruction.
The azimuthal angle reconstruction bias is also smaller than that from conventional method.
Though the polar angle reconstruction bias is larger than that from conventional method,
the bias can be calibrated.
\end{multicols}
\begin{center}
\tabcaption{ \label{direction_result}  Comparison of the tracks direction reconstruction results between two methods.}
\footnotesize
\begin{tabular*}{170mm}{c|@{\extracolsep{\fill}}cccc}
\toprule
&$\sigma_{\theta} (10^{-1}\rm{mrad})$ & $\Delta_{\theta} (10^{-3} \rm{mrad})$ & $\sigma_{\phi} (\rm{mrad})$& $\Delta_{\phi} (10^{-2}\rm{mrad})$\\\hline
Conventional method    &$1.21\pm0.01$   &   $0.60\pm0.44$   &   $5.86\pm0.02$   &   $-1.75\pm2.08$\\\hline
DNN method   &$0.98\pm0.01$   &   $4.22\pm0.35$   &   $4.61\pm0.02$   &   $-0.97\pm1.66$\\
\bottomrule
\end{tabular*}
\end{center}
\begin{multicols}{2}

\section{Summary and discussion}

This work has studied the incident track energy and direction reconstruction method of
the CEPC luminometer and presents the performance about the energy
and position resolution of the luminometer, based on the current structure and geometry design.
Firstly, we has introduced the conventional track reconstruction method presents
the track reconstruction results of it.
The conventional energy reconstruction method can not deal with the tracks falling
into the tile gap regions in energy reconstruction, which causes
about $40\%$ efficiency loss.
To solve the problems of the conventional method, we introduce a 6 layers DNN model
working on the energy reconstruction. Without discarding the gap region events,
the DNN method can
reach a better energy resolution than the conventional way.
Encouraged by the good performance of the DNN energy reconstruction method,
we extend the DNN method to
the track direction reconstruction and also get better performance
than the conventional method.
This means we can use the deep learning
method to extract original track information from the deposited energies in sensor cells,
and provide a more precise result than the conventional method.
The DNN method can be applied as benchmark reconstruction algorithm for the CEPC luminometer, and further detector optimization will rely on it. As a promising method, DNN will be used in more CEPC technique studies such as reconstruction of tracks in other sub-detectors, trigger system, fast simulation, physics analysis, and so on.

\acknowledgments{The numerical calculations in this paper have been done on the supercomputing system in the Supercomputing Center of Wuhan University.}

\end{multicols}

\vspace{-1mm}
\centerline{\rule{80mm}{0.1pt}}
\vspace{2mm}

\begin{multicols}{2}

\end{multicols}

\clearpage

\begin{thebibliography}{90}

\vspace{3mm}

%
%



\bibitem{CEPC_PreCDR}
  CEPC-SPPC Study Group,
  ``CEPC-SPPC Preliminary Conceptual Design Report. 1. Physics and Detector,''
  \href{http://inspirehep.net/record/1395734}
  {IHEP-CEPC-DR-2015-01, IHEP-TH-2015-01, IHEP-EP-2015-01.}

\bibitem{Lcal_geo}
J. Blocki, W. Daniluk, E. Kielar, J.Kotula, A.Moszczy$\acute{n}$ski, K.Oliwa, B. Pawlik, W. Wierba, L. Zawiejski and J. Aguilar on behalf of the FCAL collaboration,
``Lumi- Cal new mechanical structure'',
\href{https://www.eudet.org/e26/e28/e42441/index_eng.html}
{https://www.eudet.org/e26 /e28/e42441/index\_eng.html}

\bibitem{ref_DNN}
Y. LeCun, Y. Bengio, and G.Hinton,
\href{https://www.nature.com/articles/nature14539}
{Nature, 2015, {\bf 521}: 436---444}

\bibitem{DNN_work1}
S. Delaquis {\it et al.}
\href{http://stacks.iop.org/1748-0221/13/i=08/a=P08023}
{JINST, 2018, {\bf 13}: P08023}

\bibitem{DNN_work2}
R. Acciarri {\it et al.}
\href{https://doi.org/10.1088/1748-0221/12/03/P03011}
{JINST, 2017, {\bf 12}: P03011}

\bibitem{DNN_work3}
P. Baldi, P. Sadowski and D. Whiteson,
\href{https://www.nature.com/articles/ncomms5308}
{Nature Commun. 2014, {\bf 5}: 4308}

\bibitem{DNN_work4}
A. Aurisano {\it et al.}
\href{https://doi.org/10.1088/1748-0221/11/09/P09001}
{JINST, 2016, {\bf 11}: P09001}

\bibitem{DNN_work5}
J. Renner {\it et al.}
\href{https://doi.org/10.1088/1748-0221/12/01/T01004}
{JINST, 2017, {\bf 12}: T01004}


\bibitem{DNN_work6}
 H.~Qiao, C.~Lu, X.~Chen, K.~Han, X.~Ji and S.~Wang,
  \href{https://doi.org/10.1007/s11433-018-9233-5}
  {Sci.\ China Phys.\ Mech.\ Astron.\ 2018, {\bf 61}: 101007}
\bibitem{DNN_work7}
  M.~Paganini, L.~de Oliveira and B.~Nachman,
   \href{https://doi.org/10.1103/PhysRevLett.120.042003}
  {Phys.\ Rev.\ Lett.\ 2018,\ {\bf 120}: 042003}



\bibitem{geant4} S.~Agostinelli {\it et al.} ({\sc GEANT4} Collaboration),
\href{https://doi.org/10.1016/S0168-9002(03)01368-8}
{Nucl.\ Instrum.\ Meth.\ A\ 2003,\ {\bf 506}: 250---303}

\bibitem{mokka}
P.Mora de Fretas and H. Videau,
``Detector simulation with MOKKA/GANT4: Present and future''
\href{http://ilcsoft.desy.de/portal/software_packages/mokka/marlin.index_eng.html}
{http://ilcsoft. desy.de/portal/software\_packages/mokka/marlin.index\_eng.html}
\bibitem{Clustering_method}
  I.~Sadeh,
  ``Luminosity Measurement at the International Linear Collider,''
   \href{https://arxiv.org/abs/1010.5992}
  {[arXiv:1010.5992 [physics.acc-ph]]}

\bibitem{Marlin}
``MARLIN - a C++ software framework for ILC software."
\href{http://ilcsoft.desy.de/Marlin/current/doc/html/index.html}
{http://ilcsoft.desy.de/Marlin/current/doc/html/index.html}

\bibitem{adam}
D. P. Kingma and J. Ba,
``Adam, A Method for Stochastic Optimization,''
\href{https://arxiv.org/abs/1412.6980}
{2014, [arXiv:1412.6980]}.

\bibitem{keras}
F. Chollet {\it et al.}
``Keras'',
\href{https://github.com/fchollet/keras}
{https://github.com/keras-team/ keras} (2015)
\bibitem{tensorflow}
M. Abadi, {\it et al.}
``TensorFlow: Large-scale machine learning on heterogeneous systems'',
\href{https://github.com/topics/tensorflow}
{https://tensorflow.org/} (2015)
\bibitem{fabjan}
Fabjan, Christian W. and Gianotti, Fabiola,
\href{https://doi.org/10.1103/RevModPhys.75.1243}
{Rev. Mod. Phys. 2003, {\bf 75}: 1243---1286
}
\end{thebibliography}

\end{CJK*}
\end{document}